\documentclass[%
superscriptaddress,
amsmath,amssymb,
aps,
floatfix,
]{revtex4-2}

\usepackage{graphicx}
\usepackage{bm}
\usepackage{hyperref}
\usepackage{lineno}
\usepackage{multirow}
\usepackage[dvipsnames]{xcolor}

\usepackage{amsfonts}
\usepackage{soul}
\begin{document}


\title{\Large{All-optical computing with beyond 100-GHz clock rates}}

\author{Gordon H.Y. Li$^{1,*}$, Midya Parto$^{2,3,4,*}$, Jinhao Ge$^{5,*}$, Qing-Xin Ji$^{5}$, Maodong Gao$^{3, 5}$, Yan Yu$^{5}$, James Williams$^{2}$, Robert M. Gray$^{2}$, Christian R. Leefmans$^{1}$, Nicolas Englebert$^{2}$, Kerry J. Vahala$^{5}$, and Alireza Marandi$^{1,2,\dagger}$\\
\bigskip
\small{$^{1}$Department of Applied Physics, California Institute of Technology, Pasadena, CA 91125, USA\\
$^{2}$Department of Electrical Engineering, California Institute of Technology, Pasadena, CA 91125, USA\\
$^{3}$Physics and Informatics Laboratories, NTT Research, Inc., Sunnyvale, California 94085, USA\\
$^{4}$CREOL, The College of Optics and Photonics, University of Central Florida, Orlando, FL, USA\\
$^{5}$T. J. Watson Laboratory of Applied Physics, California Institute of Technology, Pasadena, California 91125, USA\\
$^{*}$These authors contributed equally\\
$^{\dagger}$marandi@caltech.edu\\}}

\begin{abstract}
\normalsize{\noindent \textbf{Abstract:} A computer's clock rate ultimately determines the minimum time between sequential operations or instructions. Despite exponential advances in electronic computer performance owing to Moore's Law and increasingly parallel system architectures, computer clock rates have remained stagnant at $\sim5~\mathrm{GHz}$ for almost two decades. This poses an intractable problem for applications requiring real-time processing or control of ultrafast information systems. Here we break this barrier by proposing and experimentally demonstrating computing based on an end-to-end and all-optical recurrent neural network harnessing the ultrafast nature of linear and nonlinear optical operations while avoiding electronic operations. The all-optical computer realizes linear operations, nonlinear functions, and memory entirely in the optical domain with $>100~\mathrm{GHz}$ clock rates. We experimentally demonstrate a prototypical task of noisy waveform classification as well as perform ultrafast in-situ analysis of the soliton states from integrated optical microresonators. We further illustrate the application of the architecture for generative artificial intelligence based on quantum fluctuations to generate images even in the absence of input optical signals. Our results highlight the potential of all-optical computing beyond what can be achieved with digital electronics by utilizing ultrafast linear, nonlinear, and memory functions and quantum fluctuations.}
\end{abstract}

\maketitle
\newpage
\section*{Introduction}
The clock rate ultimately determines the minimum time between sequential operations or instructions in a computer~\cite{hennessy2011computer}, and a computer cannot effectively process information or respond to input signals occurring on timescales faster than a single clock cycle. The evolution of computer hardware has been characterized by many major technological shifts: starting from early mechanical computers such as the Z1~\cite{bauer2009origins} with a clock rate of $1~\mathrm{Hz}$, then progressing to general purpose electronic computers constructed from vacuum tubes such as ENIAC~\cite{burks1947electronic} with a clock rate of $100~\mathrm{kHz}$, and finally maturing into today's central processing units (CPUs)~\cite{hennessy2011computer} consisting of billions of integrated silicon transistors with GHz clock rates. Each dramatic increase in clock rate throughout history has yielded countless new applications and innovations which were previously computationally infeasible. 

\begin{figure}[t]
\includegraphics[width=\linewidth]{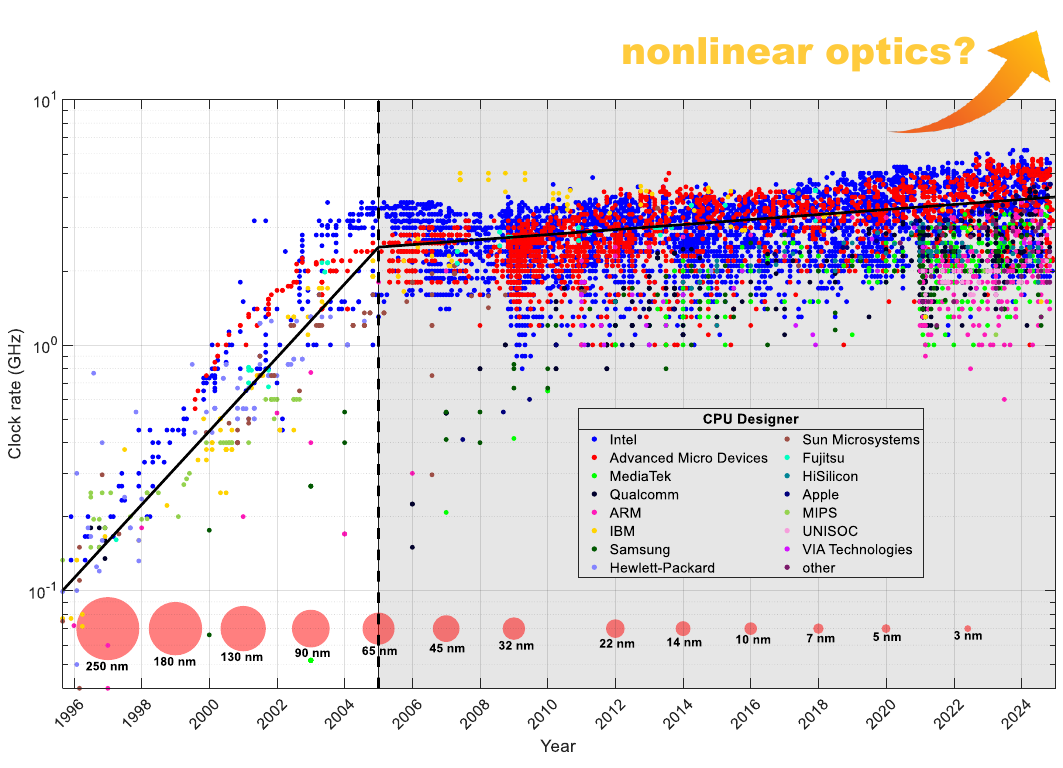}
\caption{\textbf{CPU clock rates over the past 29 years.} Each point indicates the clock rate and testing date for a different type of commercially-available CPU. The colour corresponds to the entity that designed the CPU, with some prominent CPU designers listed in the legend. Red circle and text bottom insets represent increasingly advanced semiconductor process nodes and approximately when they were introduced. Prior to 2005, clock rates increased exponentially and commensurately with Moore's Law. However, clock rates have stagnated and only increased incrementally since 2005.}
\label{fig:1}
\end{figure}

Modern CPU clock rates have stagnated at $\sim5~\mathrm{GHz}$ since circa 2005 as shown in Fig.~\ref{fig:1}. Prior to 2005, CPU clock rates increased commensurately with Moore's Law~\cite{schaller1997moore}. This abrupt change is mainly due to the breakdown of Dennard scaling~\cite{dennard1974design} for transistors at the device level and the increasing prevalence of the von-Neumann bottleneck~\cite{backus1978can} at the system level, which prompted CPU designers to abandon further significant increases in clock rate. Indeed, recent gains in computer performance can be largely attributed to the introduction of multi-core and other highly parallel computer architectures. Although the clock rate is not a directly comparable measure of computing speed between different families of computer processors since the instruction sets and operations during each clock cycle may differ, it remains clear that the limited clock rates of electronic computers preclude real-time processing or control of emerging ultrafast information systems at picosecond or faster timescales. This highlights a unique opportunity for optical computing, particularly where all the computational operations, i.e. linear and nonlinear functions, as well as the memory, are realized in the optical domain. Such an all-optical computing platform has been challenging to implement in a scalable and programmable fashion especially because of the limitations with the nonlinear functions and all-optical memory~\cite{mcmahon2023physics}.

Optical computers have experienced a resurgence in recent years as application-specific hardware for both linear operations~\cite{heinz1970matrix,xu202111,feldmann2021parallel} and nonlinear functions~\cite{li2023all,guo2022femtojoule,mourgias2019all} in deep learning~\cite{shen2017deep,ashtiani2022chip,bandyopadhyay2024single}, neuromorphic computing~\cite{feldmann2019all,tait2017neuromorphic,prucnal2017neuromorphic,brunner2025roadmap}, and combinatorial optimization~\cite{marandi2014network,inagaki2016coherent,honjo2021100} workloads. However, previous approaches mainly prioritized energy-efficient, high-throughput, or parallel processing and still relied on digital processors or optoelectronics to perform intermediate steps of the computation, hence were ultimately bottlenecked by electronic response times. Additionally, several of the previous all-optical approaches~\cite{marandi2014network,parto2020realizing,lin2018all,duport2012all,dejonckheere2014all,zuo2019all} suffered from one or a combination of (i) lacking crucial operations such as nonlinearity and/or memory, (ii) lack of programmability, and (iii) utilization of slow nonlinearities.

An all-optical and programmable computer that can exceed the clock rates of current electronic computers is lacking. In this work, we propose and experimentally demonstrate all-optical computing at high clock rates beyond the limitations of electronic computers by combining ultrafast nonlinear optics for nonlinear operations, interference for linear operations, and active cavities for optical memories. Ultrafast nonlinear optics provides two unique advantages over previous approaches: (i) the near-instantaneous response time of the parametric $\chi^{(2)}$ optical nonlinearity is orders-of-magnitude faster than electronic nonlinearities, and (ii) the ability to generate ultrashort laser pulses, which allows for time-multiplexing and higher single-channel data-encoding rates compared to electronics or continuous-wave light. We provide several experimental examples of computing tasks including noisy waveform classification with clock rates $>100~\mathrm{GHz}$, in-situ processing of native ultrafast optical input signals, time-series forecasting, and all-optical image generation seeded from quantum noise.

\section*{Results}
\subsection*{All-optical computer architecture}
\begin{figure}[b]
\includegraphics[width=\linewidth]{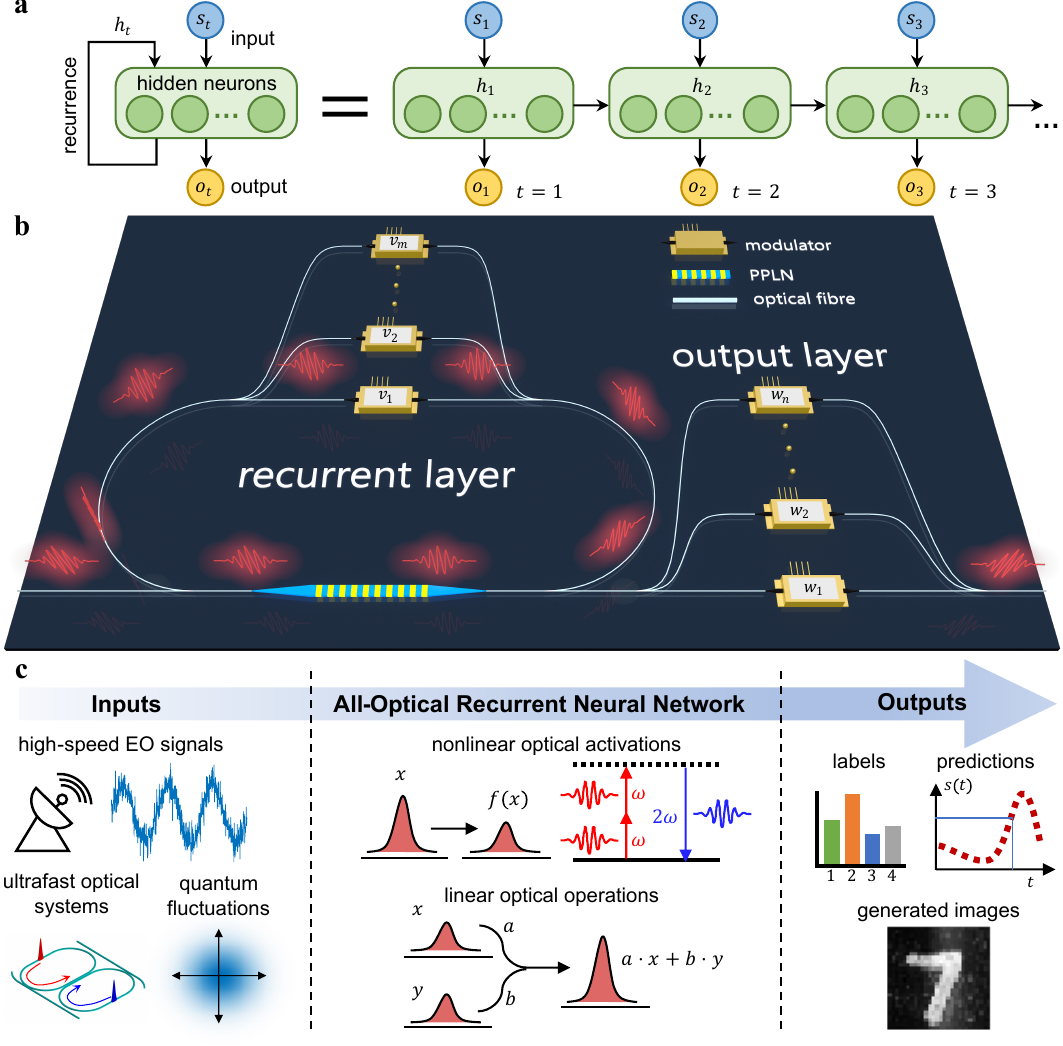}
\caption{\textbf{All-optical computer architecture.} (a) A general recurrent neural network (RNN) consists of an input layer, hidden layer with recurrent connections, and output layer. (b) Schematic of the end-to-end all-optical recurrent neural network (AO-RNN) architecture based on a time-multiplexed photonic network with (c) ultrafast optical inputs undergoing optical feedback and recurrent connections, nonlinear optical activations, and linear optical operations to produce optical outputs.}
\label{fig:2}
\end{figure}
The optical computer architecture is based on a recurrent neural network (RNN) as shown in Fig.~\ref{fig:2}a. An RNN contains an input layer, hidden recurrent layer, and output layer~\cite{medsker1999recurrent}. Unlike purely feed-forward architectures, this kind of driven dynamical system is well-suited for temporal or sequential information processing due to the inherent memory endowed by recurrent neuron states that are propagated between successive time steps. We also note that RNNs are Turing-complete~\cite{siegelmann1992computational}. Compared to digital or von-Neumann computer architectures, the RNN architecture lends itself more naturally to ultrafast optics since it is inherently analog and utilizes a dynamical memory instead of non-volatile memory elements. We construct an experimental proof-of-concept for an all-optical recurrent neural network (AO-RNN) using off-the-shelf optical fibre components, with operating wavelength of $\lambda\approx1.55~\mathrm{\mu m}$, as shown in Fig.~\ref{fig:2}b. The AO-RNN is based on a time-multiplexed photonic network~\cite{marandi2014network,leefmans2022topological,bai2023photonic} in which information and input data sequences $\{s_{t}\}$ are encoded onto the coherent amplitude of ultrashort laser pulses. Therefore, the effective clock rate of the AO-RNN is equivalent to the laser pulse repetition rate $f_{c}$. We utilized different kinds of optical frequency combs~\cite{fortier201920} including mode-locked lasers, electro-optic frequency combs, and optical microcombs to generate input signals for different tasks. Recurrent connections between time steps are performed using an active optical cavity, which acts as an optical feedback loop. The optical cavity also contains intra-cavity linear couplings implemented using a multi-arm Mach-Zehnder interferometer in which the coupling weights are encoded using electro-optic amplitude modulators in each arm. The specific network topology of the recurrent layer is determined by the lengths of the optical delay lines in each arm, so the temporal delays $T_{m}$ should be an integer multiple of the clock period $1/f_{c}$. In-line nonlinear activation functions are performed using a reverse-proton exchange periodically-poled lithium niobate (PPLN) waveguide~
\cite{langrock2007fiber}. The PPLN enables strong $\chi^{(2)}$ nonlinear optical processes such as pump-depleted second-harmonic generation at low pulse energies, which results in a sigmoid-like input-output function for the pulse amplitude at the fundamental harmonic~\cite{li2024deep}. Finally, the linear output layer is performed using another multi-arm Mach-Zehnder interferometer with weights encoded by electro-optic modulators and connections determined by temporal delay lines. Therefore, the AO-RNN is an end-to-end and fully-analog optical computer that accepts ultrafast optical inputs and produces optical outputs through a combination of linear operations, nonlinear activations, and memory feedback entirely in the optical domain. 

\subsection*{Noisy waveform classification}
\begin{figure}[b]
\includegraphics[width=\linewidth]{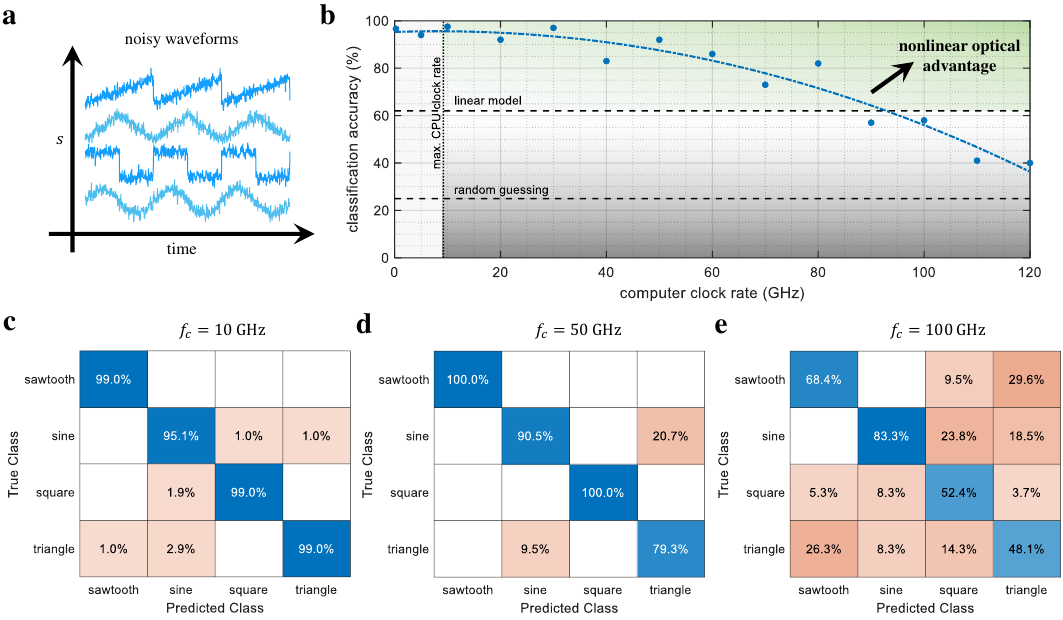}
\caption{\textbf{All-optical noisy waveform classification.} (a) Temporal seqeunces of sawtooth, triangle, square, and sine waves corrupted with noise are classified by the AO-RNN. (b) The measured classification accuracy (blue dots) generally decreases as the clock rate $f_{c}$ increases. The bottom horizontal dashed line corresponds to random guessing accuracy of $25\%$, the upper horizontal dashed line corresponds to the accuracy of a purely linear model, and the vertical dashed line indicates the maximum achievable clock rates by digital electronic computers. Therefore, a nonlinear optical computational advantage is achieved in the green-shaded region. The (column-normalized) confusion matrices are shown for clock rates of (c) $f_{c}=10~\mathrm{GHz}$, (d) $f_{c}=50~\mathrm{GHz}$, and (e) $f_{c}=100~\mathrm{GHz}$.}
\label{fig:3}
\end{figure}
To test the maximum possible clock rate $f_{c}$ of the experimental AO-RNN, we trained it (see Materials and Methods) to perform the prototypical task of noisy waveform classification. We consider four classes of temporal waveforms: sawtooth, triangle, square, and sine. The waveforms have the same period and duration, but are also corrupted by some random noise as shown in Fig.~\ref{fig:3}a. The waveform samples are fed sequentially as a single-channel input into the AO-RNN, and the task is to classify each waveform into the correct class. Therefore, this task is inherently single-threaded and cannot necessarily be accelerated through parallel processing. The AO-RNN produces a single-channel optical output pulse for each optical input pulse. To assign a single class label for verification purposes, the output pulses are photodetected, which effectively acts as a low-pass filter performing temporal global average pooling. The average optical output is compared against threshold decision boundaries (Materials and Methods) in digital postprocessing to assign a final class label. Alternatively, if multiple output channels are available, then a conventional softmax classification can also be applied. Note that the digital postprocessing required to assign class labels in this case only occurs after the entire duration of the waveform sequence, hence it only needs to be performed at a much lower rate compared to the clock rate $f_{c}$ of the AO-RNN and does not bottleneck the computation. In principle, the output optical pulses containing information about the classification result could be used as the input to a second optical computer stage for further optical processing if desired.  

The classification accuracy of the AO-RNN generally decreases as the clock rate $f_{c}$ increases as shown in Fig.~\ref{fig:3}b. It achieves a peak classification accuracy of $97.5\%$ at a clock rate of $f_{c}=10~\mathrm{GHz}$ (Fig.~\ref{fig:3}c), which exceeds the clock rate of commercially-available CPUs. The classification accuracy decreases to $92\%$ at a clock rate of $f_{c}=50~\mathrm{GHz}$ (Fig.~\ref{fig:3}d) as the AO-RNN begins to confuse some sine/triangle waveforms. The classification accuracy further decreases to $58\%$ at a clock rate of $f_{c}=100~\mathrm{GHz}$ (Fig.~\ref{fig:3}e) as the AO-RNN also confuses square/sine and triangle/sawtooth waveforms. Even at the maximum tested clock rate of $f_{c}=120~\mathrm{GHz}$, the AO-RNN still achieves a classification accuracy significantly higher than random guessing. Repeating the same task using a purely linear model (i.e. without nonlinear activation function) with $f_{c}=10~\mathrm{GHz}$ achieves a classification accuracy of only $62\%$, which confirms that the PPLN optical nonlinearity is important for achieving a high classification accuracy. We designate an AO-RNN with both a clock rate higher than any electronic CPU and classification accuracy higher than a linear model as exhibiting a ``\textit{nonlinear optical advantage}''.

We have identified several reasons explaining why the classification accuracy decreases as the clock rate $f_{c}$ increases. First, for testing purposes, we generated the input signals electro-optically using optical time-interleaving techniques (Supplementary Information Section V), which becomes increasingly more difficult as $f_{c}$ increases. The AO-RNN relies on coherent interference for performing linear operations and memory feedback, hence the relative phase and temporal separation of coherent laser pulses is critically important. Therefore, the computation fidelity decreases as the phase noise and timing jitter in the input signal increases. Second, although the $\chi^{(2)}$ optical nonlinearity is of the ultrafast variety and near-instantaneous, the PPLN possesses a finite phase-matching bandwidth $\sim100~\mathrm{GHz}$. Therefore, the effectiveness of the nonlinear activation function , and hence classification accuracy, begins to degrade when $f_{c}$ exceeds the phase-matching bandwidth. Finally, neighbouring pulses begin to overlap temporally and experience undesirable cross-talk when the clock period $1/f_{c}$ becomes comparable to the pulse duration.  

\subsection*{Native ultrafast optical signals}
Many of the above-mentioned issues can be avoided entirely if the input data occurs natively in the optical domain, thus not requiring electro-optic or optoelectronic conversions for generating input signals. Fortunately, the study of ultrafast optical science is rife with examples of optical signals that possess a high degree of coherence and occur on timescales that are too fast for real-time processing or control by electronic computers. Here we introduce the AO-RNN as an in-situ tool for ultrafast optical science, which can potentially enable new functionalities that are infeasible using existing experimental techniques. 

\begin{figure}[t]
\includegraphics[width=\linewidth]{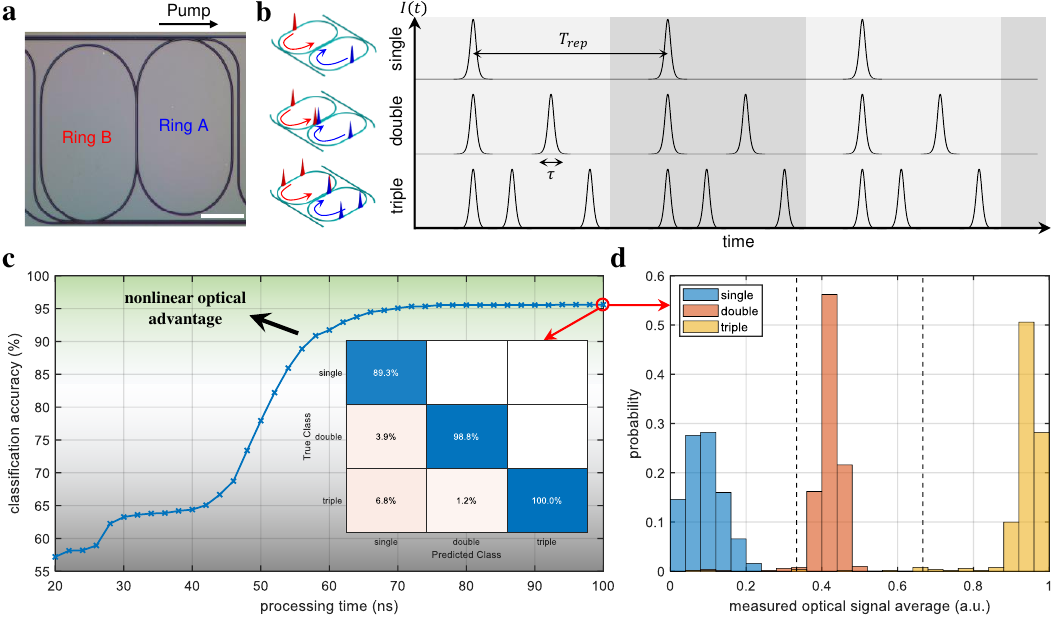}
\caption{\textbf{Microcomb bipartite soliton state classification.} (a) Micrograph of a $\mathrm{Si_{3}N_{4}}$ integrated coupled microresonator (white scale bar: $1~\mathrm{mm}$). (b) Bipartite bright soliton states with single, double, or triple pulse pairs. (c) Classification accuracy of soliton states with different AO-RNN processing times. Inset: confusion matrix for a processing time of $100~\mathrm{ns}$. (d) Histogram and decision boundaries for measured optical output averages with a processing time of $100~\mathrm{ns}$.}
\label{fig:4}
\end{figure}

For example, Fig.~\ref{fig:4}a shows ultra-low-loss $\mathrm{Si_{3}N_{4}}$ integrated coupled microresonators~\cite{yuan2023soliton}. This configuration can be operated as an optical frequency comb or ``microcomb" through the balance of $\chi^{(3)}$ optical nonlinearity and dispersion, giving rise to bipartite bright-soliton states. Microcombs have attracted immense research interest due to their technological importance in a wide range of applications including optical computing, LIDAR, dual-comb spectroscopy, low-noise microwave synthesis, optical metrology, and astrocombs~\cite{xu202111,feldmann2021parallel,doi:10.1126/science.aan8083}. The $\mathrm{Si_{3}N_{4}}$ microcomb can produce different soliton states including a single soliton pulse pair, double soliton pulse pair, or triple soliton pulse pair (Supplementary Information Fig.~S6). The output optical signal in the time-domain is a periodic waveform composed of sub-picosecond pulses with repetition rate of $f_{rep}\approx19.97~\mathrm{GHz}$ as shown in Fig.~\ref{fig:4}b. We consider the task of classifying bipartite soliton states (single, double, or triple) given that the optical waveforms have the same average power, repetition rate, centre wavelength, polarization, and spectral bandwidth. This is challenging to do in real-time since the optical waveforms are too fast for direct photodetection. Moreover, conventional optical measurement techniques using spectrometers or autocorrelation often require scanning elements at millisecond or slower timescales. Other ultrafast characterization methods such as time-stretching~\cite{herink2017real,foster2008silicon,goda2013dispersive} struggle with the combination of both high-repetition rate and high-duty cycle, which is needed to distinguish the bipartite soliton states. Existing single-shot~\cite{kane1993single} or few-shot~\cite{yi2018imaging} methods suffer from low frame update rates and/or require extensive postprocessing algorithms to extract the useful information. Additionally, our all-optical computing architecture can be realized as a feedback mechanism for controlling the soliton state generation and/or ultrafast sensing schemes based on similar soliton dynamics~\cite{gray2024quadratic}.

The AO-RNN can classify bipartite soliton states with varying amounts of processing time given by the length of the input optical waveform as shown in Fig.~\ref{fig:4}c. The final class label is assigned in the same way as for noisy waveform classification. The AO-RNN achieves a high classification accuracy of $95.6\%$ for processing times shorter than $100~\mathrm{ns}$. The minimum processing time or latency is limited by the main cavity roundtrip light propagation time in the AO-RNN, which is $\sim24~\mathrm{ns}$. We observe that multiple cavity roundtrips are necessary to ensure high classification accuracy. This latency is a consequence of using relatively long optical-fibre components/connectors, and can be drastically reduced by direct splicing or instead using a corresponding integrated photonic platform with sub-nanosecond roundtrip times. Therefore, it is possible to use the AO-RNN as an in-situ tool for near-real-time and few-shot classification of optical soliton states, which can enable faster measurement and feedback control loops. 

\subsection*{Time-series forecasting}
\begin{figure}[t]
\includegraphics[width=\linewidth]{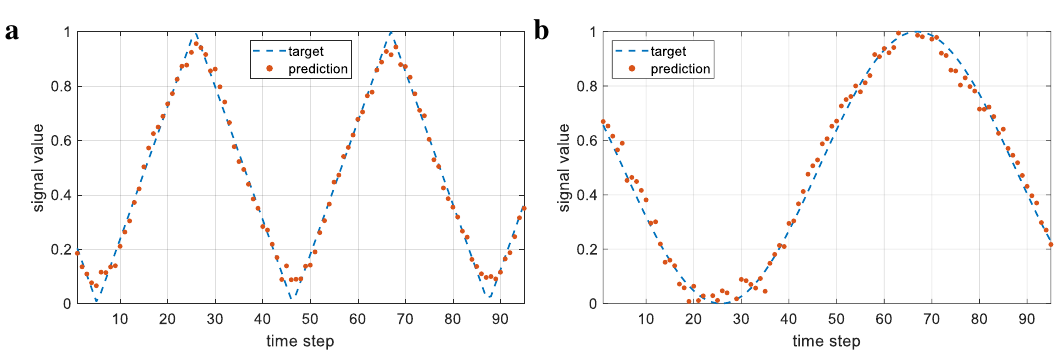}
\caption{\textbf{Temporal waveform prediction.} Predicted (dots) one-step-ahead values for target (dashed lines) (a) triangle and (b) sine input time-series.}
\label{fig:5}
\end{figure}

The previous two example tasks were classification tasks. We further show that the AO-RNN can also be trained to perform regression tasks such as time-series forecasting with a high clock rate. It is desirable to have faster regression methods to make real-time decisions in many applications including quantitative finance~\cite{gomber2013high}, experimental particle physics~\cite{gligorov2013efficient}, and optical signal processing~\cite{willner2013all}. In this task, samples of a time-series are encoded onto the coherent amplitude of optical pulses and inputted one at a time into the AO-RNN. The task is to predict the next value in the time-series given the current input value and past inputted values. The corresponding output pulse should have an amplitude representing the predicted value of the next time step sample. 

We show two examples of time-series forecasting for triangle and sine waveforms in Fig.~\ref{fig:5}. The AO-RNN predictions show close agreement with the target values for these simple waveforms, and achieves normalized mean square error (NMSE) as low as 0.0144 and 0.0094, respectively, up to a clock rate of $f_{c}=10~\mathrm{GHz}$. Unlike for classification tasks in which the speed of the output measurement can be amortized over the entire duration of the time-series, regression tasks require single-shot and rapid output measurements. In this case, our ability to test and verify the output predictions is limited by the maximum bandwidth of our photodetector that is $\sim25~\mathrm{GHz}$. Although the AO-RNN can operate with far higher clock rates, as evidenced by the noisy waveform classification task, we were experimentally limited in accurately measuring the real-time pulse-to-pulse output values beyond a clock rate of $f_{c}=10~\mathrm{GHz}$. This may be improved by using a faster photodetector up to $\sim100~\mathrm{GHz}$ and better data acquisition tools. Nevertheless, this is not a fundamental limitation of the AO-RNN, but rather a constraint of our current output measurement techniques. If an output remaining in the optical domain is sufficient, then the clock rate may be much higher since the optical outputs can in principle directly serve as optical inputs for another optical computer or actuator.  

\subsection*{Image generation seeded from quantum fluctuations}
Finally, we demonstrate an example of a generative task in which the AO-RNN can use quantum fluctuations as the seed to generate images in the absence of any input optical signals~\cite{choi2024photonic,roques2023biasing}. The central problem for generative models is to learn a complicated unknown target distribution from which samples (e.g. images) are available, and then to use the model to efficiently generate new samples from the target distribution. We take inspiration from recent advances in generative artificial intelligence based on diffusion~\cite{ho2020denoising} and flow-based~\cite{lipman2022flow} models. In this case, a simple known distribution (typically a standard Gaussian distribution) is continuously perturbed to match the unknown target distribution. Then, the learned mapping can be applied to an initial random sample from the simple distribution to generate a new sample from the target distribution. 

\begin{figure}[t]
\includegraphics[width=\linewidth]{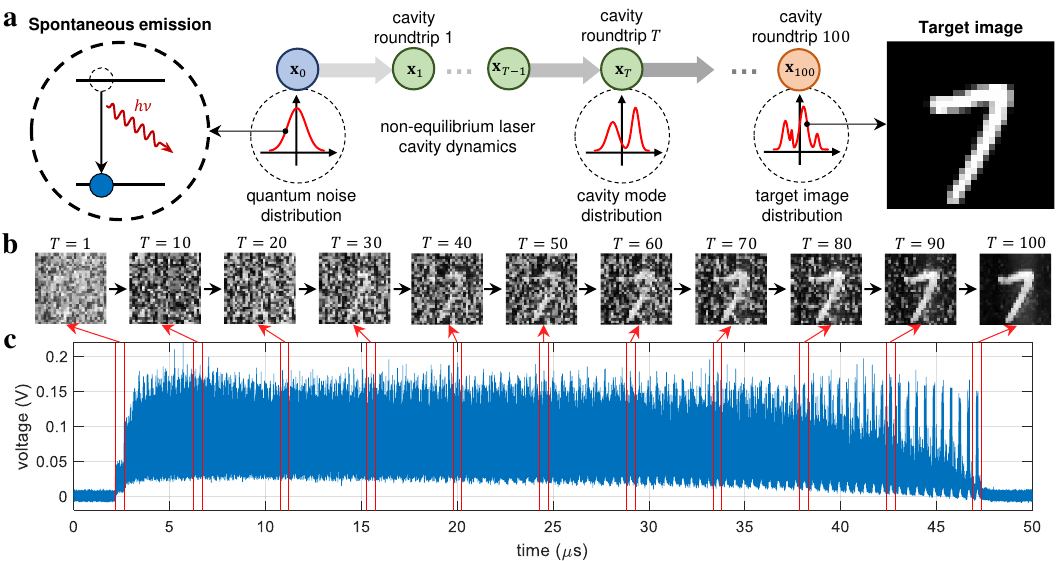}
\caption{\textbf{Quantum all-optical image generation.} (a) An all-optical image generator maps a quantum noise distribution from spontaneous emission into an unknown target image distribution of MNIST handwritten digits using highly non-equilibrium laser cavity dynamics. (b) Example images after every 10 cavity roundtrips and (c) measured cavity output time trace up to $T=100$ roundtrips representing a generated training sample of a seven.}
\label{fig:6}
\end{figure}

We trained the AO-RNN using the MNIST handwritten digits dataset~\cite{deng2012mnist} to generate $28\times28$ greyscale images of the class ``seven'' as shown in Fig.~\ref{fig:6}a. For this task, we ignore the input and output layers of the AO-RNN and only directly utilize the recurrent layer. The main cavity contains an optical amplifier that acts as a programmable gain/loss mechanism. The optical amplifier also supplies quantum noise in the form of spontaneous emission, which approximately follows a Gaussian distribution (Supplementary Information Fig.~S9). Therefore, the AO-RNN can be interpreted as a highly nonlinear laser cavity. In conventional lasers, spontaneous emission in the gain medium serves as the spark to initiate stimulated emission, whereupon population inversion and gain-clamping after the laser threshold (i.e. gain equals loss) is exceeded will lead to a steady-state laser emission~\cite{milonni2010laser}. Here we modulate the gain and intra-cavity connection weights at a clock rate $f_{c}$ that is much faster compared to the cavity roundtrip time. Therefore, the lasing dynamics are also highly non-equilibrium. We define cavity modes by virtually splitting the cavity into equally-spaced time bins. Each time bin corresponds to a different pixel location in the image and the average power contained in each time bin encodes for the pixel greyscale intensity. Upon initially turning on, the AO-RNN starts from spontaneous emission in the optical amplifier and then gradually generates a macroscopic image after $T=100$ cavity roundtrips by effectively controlling which cavity modes temporarily go above/below lasing threshold.

An example time trace of the cavity dynamics and resultant image generated of a sample during the training process is shown in Fig.~\ref{fig:6}b, c. Examples of new images (i.e. not part of the original MNIST dataset) of sevens generated after the training process and seeded by quantum noise are shown in Supplementary Information Fig.~S10. We note that our simple generative AO-RNN lacks model expressiveness and struggled to generate high-quality images of multiple different digit classes or learn more complicated image distributions beyond the MNIST dataset. We believe that the quality and diversity of generated sevens from this simple proof-of-concept model is promising, and better model expressiveness can be achieved through exploring more complicated AO-RNN architectures. 

\section*{Discussion}
There are a few key timescales limiting the maximum allowable clock rate of the AO-RNN. We opted to use off-the-shelf optical fibre components for simplicity, however, this was not ideal for optimal computer performance. For example, the PPLN used for nonlinear activations based on a weakly-guiding reverse-proton exchange waveguide~\cite{langrock2007fiber} has a relatively small phase-matching bandwidth of $\sim100~\mathrm{GHz}$. However, we previously demonstrated higher-performance nonlinear activation functions using thin-film lithium niobate (TFLN) with maximum allowable clock rates $>13~\mathrm{THz}$~\cite{li2023all}. Therefore, the AO-RNN could greatly benefit from on-chip integration using TFLN. Furthermore, other devices demonstrated in TFLN, such as high-speed electro-optic modulators with $>100~\mathrm{GHz}$ bandwidth~\cite{wang2018integrated} or all-optical switches with speeds $>10~\mathrm{THz}$~\cite{guo2022femtojoule} can enable faster real-time updating of weights. The longest duration laser pulse width used in our experiments was $\sim5~\mathrm{ps}$, which limits the maximum allowable clock rates since pulses will begin overlapping in time and suffering from undesirable cross-talk beyond $\sim200~\mathrm{GHz}$ clock rates. This issue can be overcome by using even shorter laser pulses, for example, few-cycle pulses generated using nonlinear optical pulse compression~\cite{gray2024soliton} in TFLN. Finally, implementing the main cavity in the AO-RNN using an integrated optical parametric oscillator~\cite{roy2023visible,ledezma2023octave,gray2024large} in TFLN could also drastically reduce the overall latency given by the light propagation time through the network. 

The conventional definition of clock rate may not be directly applicable to some kinds of asynchronous or clock-less analog computers~\cite{ashtiani2022chip,zhou2023ultrafast}. Here we use the concept of clock period more generally to mean the minimum time between successive computer operations. This concept can thus still apply to clock-less processors since there is a characteristic physical timescale associated with each nonlinear device (electronic, optoelectronic, optical, or otherwise). The key advantage of the AO-RNN is that it implements all linear operations, nonlinear activations, and memory feedback directly in the optical domain so that all operations can simultaneously benefit from exceptionally high clock rates without an electronic bottleneck for some parts of the computation. During each clock period, the experimental AO-RNN performs a maximum of 8 operations (1 nonlinear activation and 7 multiply-accumulate operations). This number can be further increased by including more optical delay lines and modulators, as well as utilizing wavelength and/or spatial multiplexing techniques~\cite{bai2023photonic}. We believe that the most useful near-term applications for this kind of ultrafast optical computer will be those in which the input signal occurs natively in the optical domain, hence bypassing the need for electro-optic input signal generation. Some prime examples include in-situ analysis of ultrafast imaging and spectroscopy data~\cite{wang2020single,maiuri2019ultrafast}, optical signal processing for high-speed coherent telecommunications systems~\cite{willner2013all,kikuchi2015fundamentals}, and precision ranging or LIDAR using femtosecond lasers~\cite{suh2018soliton,swann2006frequency}. 

In conclusion, we have harnessed ultrafast nonlinear optics to build a new kind of end-to-end optical computer that can surpass the limited clock rates of existing digital electronic computers. The proof-of-concept experimental results for a combination of classification, regression, and generative computational tasks demonstrate that the AO-RNN can achieve clock rates $>100~\mathrm{GHz}$, which is more than an order-of-magnitude improvement over current digital computers. This work highlights a new regime for ultrafast optical computing, enabling nascent applications requiring real-time information processing and feedback control at picosecond timescales. 

\section*{Materials and Methods}
\subsection*{Experimental setup}
A detailed schematic of the experimental setup for the AO-RNN is shown in Supplementary Information Fig.~S1. Wherever possible, we used commercially-available, single-mode polarization-maintaining (PM) optical fibre components with centre operating wavelength of $\lambda\approx1.55~\mathrm{\mu m}$. The recurrent layer consists of a main cavity and contains an intra-cavity Mach-Zehnder interferometer with two arms. The relative temporal delay $T_{0}$ between the arms of the Mach-Zehnder interferometer determines the connection topology, and is a hyper-parameter of the AO-RNN to be chosen depending on the task. For proper operation, $T_{0}$ is chosen to be an integer multiple of the clock period $1/f_{c}$. The relative delay is fine-tuned using a homemade free-space delay stage since it is difficult to cut/splice optical fibers to precisely match the desired length. The approximate temporal delay is set by propagating a single reference laser pulse and manually moving the stage. Then, fine-tuning is done by a high-precision linear micrometer stage to maximize the pulse temporal overlap by observing the coherent interference fringe visibility. The weights are set using electro-optic amplitude modulators (IXBlue MXAN-LN-10) in each arm of the Mach-Zehnder interferometer. Each modulator accepts a constant bias voltage input to set the operating point, and also a high-speed RF voltage input to rapidly modulate the amplitude of optical signals. Bias voltages are controlled using a ADC/DAC (National Instruments 782258) and RF waveforms up to $12~\mathrm{GSa/s}$ are generated using an arbitrary waveform generator (Keysight M8190A). The voltages are the learnable weight parameters of the AO-RNN. We only trained the bias voltages for the noisy waveform classification, optical soliton classification, and time-series forecasting tasks. We set the bias point to ``closed'' for the image generation task and trained the RF voltages at a $10~\mathrm{GSa/s}$ update rate. The main cavity additionally contains a fibre-coupled PPLN waveguide~\cite{langrock2007fiber} with a wavelength-division multiplexer on both the input/output to separate the fundamental harmonic ($\lambda\approx1.55~\mathrm{\mu m}$) and second harmonic ($\lambda\approx0.775~\mathrm{\mu m}$). The second harmonic signal is only used to monitor the phase-matching of the PPLN, and is not used as part of the AO-RNN computation. The phase-matching of the PPLN is adjusted using a heater stage and thermocouple controller (Omega CSC32), with an optimal operating temperature around $\sim51.5^{\circ}\mathrm{C}$. The maximum average optical power in the main cavity must be $<100~\mathrm{mW}$ to avoid photo-induced damage to the PPLN. The PPLN was bypassed for the noisy waveform classification task using a purely linear model. A booster optical amplifier (Thorlabs S9FC1004P) was used to set the overall gain/loss in the main cavity for the image generation task. The amplifier was bypassed for the noisy waveform classification, optical soliton classification, and time-series forecasting tasks since we desired to have a fading memory property for these tasks. The main cavity roundtrip light propagation time was $\sim24~\mathrm{ns}$ for the noisy waveform classification, optical soliton classification, and time-series forecasting tasks. The main cavity roundtrip light propagation time was $\sim453~\mathrm{ns}$ for the image generation task since we required more time bins for a sufficient number of pixels to generate MNIST images. For the image generation task, the output of the recurrent layer was directly detected since the output layer is not used. For the other tasks requiring an output layer, the optical output of the recurrent layer is amplified using an erbium-doped fibre amplifier (Thorlabs EDFA100S) and filtered through a $200~\mathrm{GHz}$ band-pass filter to minimize the amplified spontaneous emission noise before entering the output layer. The output layer consists of a four-arm Mach-Zehnder interferometer of similar construction as the one in the recurrent layer. Three arms encode the output layer weights, and the fourth arm is used as an optional constant optical bias. The relative delays, $T_{1}\approx0.3~\mathrm{ns}$ and $T_{2}\approx1.6~\mathrm{ns}$, are hyper-parameters of the AO-RNN to be chosen depending on the task. Only the constant bias voltages were trained for the output layer modulators due to the limited number of available high-speed arbitrary waveform generator channels. Optical outputs are detected using a high-speed photodetector (Newport 1414) and stored on a real-time oscilloscope (Keysight MSOV334A) for post-processing if necessary. The high-speed RF and optical inputs are synchronized using a low-noise $10~\mathrm{MHz}$ reference clock. The relative delay between generated RF and measured optical signal is calibrated using the sample marker output from the arbitrary waveform generator. We use a backwards-propagating locking reference, which is tapped from the unmodulated input laser source, to perform active phase-stabilization of the temporal delay lines. The recurrent layer and output layer each contain independent backwards-propagating optical locking loops based on a Pound-Drever-Hall scheme~\cite{black2001introduction}. Slow photodetctors (Newport 2053) are used to measure the backwards-propagating locking signals. The electronic locking signals are input to proportional-integral derivative controllers (Red Pitaya STEMlab 125-14) and outputs are amplified (Thorlabs MDT693B) to drive fibre phase-shifters (General Photonics FPS-002-L) that actively stabilize the relative phase in each optical delay line. 

\subsection*{Input signals} 
We used a variety of different optical frequency combs to generate optical input signals for the experimental tasks. External optical signals input to the AO-RNN are gated using an electro-optic amplitude modulator, which is biased closed so that signals only enter the AO-RNN when desired to mark the start of a computation. For noisy waveform classification, the input waveforms were generated electro-optically using a high-speed arbitrary waveform generator and electro-optic amplitude modulators. The laser source for clock rates $f_{c}\in[250~\mathrm{MHz},5~\mathrm{GHz}]$ was a fibre mode-locked laser (MenloSystems FC1500-250-WG) and the laser source for clock rates $f_{c}\in[10~\mathrm{GHz},120~\mathrm{GHz}]$ was a homemade electro-optic frequency comb (Supplementary Information Section III). We can perform real-time input generation up to clock rates of $f_{c}\approx48~\mathrm{GHz}$ using optical time-interleaving (Supplementary Information Section V), and offline (i.e. prepared ahead-of-time) input generation up to clock rates of $f_{c}\approx200~\mathrm{GHz}$ using an asynchronously-pumped cavity. Each noisy waveform (sawtooth, triangle, square, and sine) was 120 periods in duration with a total of 5120 equally-spaced samples. The period for all noisy waveform classes was the same, and sawtooth/square waves had a duty cycle of $1/2$. The ideal noiseless waveforms had normalized amplitudes in the range [-1,1] and the measured optical noise for the waveforms is approximately given by an additive Gaussian distribution with zero mean and standard deviation of $\sim0.158$ as shown in Supplementary Information Fig.~S2. The $\mathrm{Si_{3}N_{4}}$ coupled optical microresonators used to generate bipartite soliton states is of the same design as in Ref.~\cite{yuan2023soliton}. We maintained the average input optical power to the AO-RNN at $\sim5.3~\mathrm{mW}$ for all bipartite-soliton states by monitoring on a slow thermal power metre (Thorlabs PM20). We carefully characterized the single, double, and triple pulse pairs by separately measuring the optical spectrum (Yokogawa AQ6370C), autocorrelation (Femtochrome FR-103XL), and RF beat-note (Rhode \& Schwarz FSW85) of each state as shown in Supplementary Information Fig.~S4. The input signals used for the time-series forecasting task were also generated electro-optically in the same way as for the noisy waveform classification. The triangle wave had 44 samples per period with a duty cycle of $1/2$. The sine wave had the same sample time-spacing as the triangle wave but double the period.

\subsection*{RNN model}
The general RNN architecture is given by Eq.~\ref{eq:1}:
\begin{subequations}
 \label{eq:1}
    \begin{equation}
        h_{i}(t+1)=f_{i}\left(\sum_{j=1}^{N}W^{r}_{ij}\cdot h_{j}(t)+\sum_{j=1}^{m}W^{in}_{ij}\cdot s_{j}(t)\right)
    \end{equation}
    \begin{equation}
        o_{k}(t)=\sum_{l=1}^{N}W^{out}_{kl}\cdot h_{l}(t)
    \end{equation}
\end{subequations}
where $t\in\mathbb{N}$ is the discrete-time step (one time-step represents one clock period), $\mathbf{h}\in\mathbb{R}^{N}$ is the $N$-dimensional hidden recurrent layer activation, $\mathbf{s}\in\mathbb{R}^{m}$ is the $m$-dimensional input sequence, $\mathbf{o}\in\mathbb{R}^{n}$ is the $n$-dimensional output value, $\mathrm{W^{r}}\in\mathbb{R}^{N\times N}$ is the matrix of recurrent layer weights, $\mathrm{W^{in}}\in\mathbb{R}^{N\times m}$ is the matrix of input layer weights, $\mathrm{W^{out}}\in\mathbb{R}^{n\times N}$ is the matrix of output layer weights, and $f_{i}:\mathbb{R}\rightarrow\mathbb{R}$ is an element-wise activation function for $i=1,2,\ldots,N$ and $k=1,2,\ldots,n$. We give a simplified model of the AO-RNN, which is similar to the rotating neuron architecture proposed in Ref.~\cite{liang2022rotating}, by dividing the main cavity into equally-spaced time bins containing pulses. The hidden recurrent layer has a cyclic structure with weights:
\begin{equation}
    W^{r}_{ij}=\begin{cases}
        \alpha_{i}(t), & \mathrm{if}\quad i-j\equiv1\ (\mathrm{mod}\ N)\\
        \beta_j(t), & \mathrm{if}\quad i=i_{r}\ \mathrm{and}\ j\in T^{r}\\
        0, & \mathrm{otherwise}
    \end{cases}
\end{equation}
where $\{\alpha_{i}(t)\}$ are weights representing the loss/gain factor for the pulse propagating from time bin $i$ to time bin $(i+1)\ \mathrm{mod}\ N$. The weights $\{\beta_{j}(t)\}$ represent the intra-cavity couplings between time bin $i_{r}$ and other time bins in $T^{r}$. The set $T^{r}$ of time bin indices represents the choice of optical delay lines in the intra-cavity Mach-Zehnder interferometer. In our experimental AO-RNN, $T^{r}=\{j_{r}\}$ represents a single connection since we used a two-arm Mach-Zehnder interferometer. This cyclic structure is a special case of the fully-connected RNN model and is still Turing-complete, however, in practice may lead to reduced model expressiveness for some tasks. We used constant weights for the noisy waveform classification, optical soliton classification, and time-series forecasting tasks, and time-varying weights for the image generation task.    
Our AO-RNN has single-channel input/outputs, $m=n=1$, due to experimental constraints. In this case, the input weights are given by $\mathrm{W^{in}}=\varepsilon\mathbf{e}_{i_{0}}$ where $\mathbf{e}_{k}$ is the $k^{\mathrm{th}}$ unit vector, $i_{0}$ is the index of the time bin coupled to the input line, and $\varepsilon$ is the input coupling factor. We treat the input scaling $\varepsilon$ as a hyper-parameter that is not trained. However, it is also possible to employ more complicated input masking techniques, such as in previous time-multiplexed photonic reservoir computers~\cite{duport2012all,dejonckheere2014all}. The output weights are given by $W^{out}_{1l}=\gamma_{l}(t)$ if $l\in T^{out}$, and $0$ otherwise, where $\{\gamma_{l}(t)\}$ are determined by the output layer modulators and $T^{out}$ is the set of time bin indices representing the choice of optical delay lines in the output layer Mach-Zehnder interferometer. For the noisy waveform and optical soliton classification tasks, we additionally perform temporal global average pooling of the optical power to yield a single output value $y=\langle\lvert o(t)\rvert^{2}\rangle_{0<t\leq L}$, over the entire input sequence of length $L$, and normalize the output averages $\{y\}$ to the range $[0,1]$. The predicted class label is assigned by comparing $y$ against threshold decision boundaries $\mathbb{Z}_{q}/q$ where $q$ is the number of classes, such that $y$ belongs to class $p\in\mathbb{Z}_{q}$ if $(y\geq p/q)\land(y<(p+1)/q)$. Alternatively, the more conventional softmax classification can be used if the number of output channels in the AO-RNN equals the number of classes. For the image generation task, only the recurrent layer is used. The input sequence can be replaced by an additive noise term (Supplementary Information Fig.~S9), representing the amplified spontaneous emission from the optical amplifier, and the output samples are then equivalent to sampling from a single point in the cavity. Strictly speaking, the nonlinear laser cavity becomes a continuous-field distribution so the concept of discrete time bins is not well-defined. However, we discretize the cavity field based on the fast output measurement sampling time $\tau$. Then, during each cavity roundtrip, we assign the value $p_{v}$ for pixel $v\in\mathbb{Z}_{V}$ in an image sequence with $V$ pixels by coarse-graining the fast output samples into slower time bins with period $t^{\prime}$ such that $p_{v}=\langle\lvert o(\tau)\rvert^{2}\rangle_{vt^{\prime}\leq\tau<(v+1)t^{\prime}}$. Pixel values are rescaled to be in the range $[-1,1]$ for each cavity roundtrip. We used a time bin period of $t^{\prime}=0.4~\mathrm{ns}$ and applied symmetric zero-padding for unused pixels in each roundtrip ($\sim453~\mathrm{ns}$) since MNIST images only contain $784$ pixels. 

\subsection*{Training procedure}
We use a model-agnostic forward-only training algorithm based on the method proposed in Ref.~\cite{bandyopadhyay2024single}. For each training iteration, we perform the following steps:

1. Choose a random direction vector $\Delta\in\{+\delta,-\delta\}^{d}$ where $d$ is the number of trainable model parameters, the elements of $\Delta$ are sampled from a Bernoulli distribution $\Delta_{i}\sim B(1/2)$ for $i=1,2,\ldots,d$, and $\delta$ is the step size. 

2. Perturb the model parameters $\Theta\in\mathbb{R}^{d}$ by $\Delta$ and perform a forward-pass through the model to evaluate the loss function $\mathcal{L}(\Theta+\Delta)$.

3. Perturb the model parameters $\Theta$ in the opposite direction $-\Delta$ and perform a forward-pass through the model to evaluate the loss function $\mathcal{L}(\Theta-\Delta)$.

4. Estimate the directional derivative of the loss as:
\begin{equation}
    \nabla_{\Delta}\mathcal{L}(\Theta)\approx\frac{\mathcal{L}(\Theta+\Delta)-\mathcal{L}(\Theta-\Delta)}{2\lVert\Delta\rVert}\ .
\end{equation}

5. Update the model parameters: $\Theta\rightarrow\Theta-\eta\nabla_{\Delta}\mathcal{L}(\Theta)\Delta$ where $\eta$ is the learning rate. 

For the experimental AO-RNN, the forward-pass steps are performed directly in the optical hardware, but the other training steps are performed on a digital computer. During testing, the trained parameters are frozen and so the forward-pass inference is all-optical. The model paramters in our experimental AO-RNN correspond to electro-optic modulator voltages (both DC and RF). The half-wave voltages are $V_{\pi}\approx6~\mathrm{V}$ and we found that a perturbation step-size of $\delta=0.02~\mathrm{V}$ with learning rate of $\eta=2\times 10^{-3}$ was adequate for all our experimental tasks. For the noisy waveform and optical soliton classification tasks, we used a mean squared error loss $\mathcal{L}=\langle(y-z)^{2}\rangle$ where $z=(2p+1)/(2q)$ is the midpoint of the decision boundaries for the true class label. For the time-series forecasting task, we used a mean squared error loss $\mathcal{L}=\langle \left[o(t)-s(t+1)\right]^{2}\rangle_{t}$ for the one-step-ahead prediction. For each training sample in the image generation task, we used a diffusion process to generate intermediate target values during each cavity roundtrip $T$:
\begin{equation}
    \mathbf{z}_{T-1}=\sqrt{1-\sigma_T}\cdot\mathbf{z}_{T}+\sqrt{\sigma_{T}}\cdot\epsilon
\end{equation}
where $\epsilon\sim\mathcal{N}(0,\mathbf{I})$ is standard Gaussian noise, $\mathbf{z}_{100}$ is the ideal target from the image dataset, and $\sigma_{T}$ is a noise-variance schedule that increases linearly each from $\sigma_{100}=0.001$ to $\sigma_{1}=0.02$. We used a mean-squared error loss $\mathcal{L}=\langle\left[\mathbf{x}_{T}-\mathbf{z}_{T}\right]^{2}\rangle_{t^{\prime},T}$ taken over all roundtrips and time bins for each sample where $\mathbf{x}_{T}$ are the measured pixel values for each roundtrip $T$. The training and testing sample sizes for each task are shown in Table.~\ref{table:1}.

\setlength{\abovecaptionskip}{5pt} 
\begin{table}[h!]
\centering
\begin{tabular}{|c|c|c|c|c|c|c|c|c|} 
\hline
task & batch size & iterations & testing size\\
\hline
noisy waveform classification & 4 & 200 & 800\\
optical soliton classification & 10 & 150 & 1500 \\
time-series forecasting & 460 & 200 & 9200 \\
all-optical image generation & 12 & 1200 & - \\
\hline
\end{tabular}
\caption{\textbf{Sample sizes for experimental AO-RNN tasks.}}
\label{table:1}
\end{table}

\subsection*{CPU clock rates}
Each scatter point in Fig.~\ref{fig:1} represents the clock rate and testing date for a different commericially-available CPU. The data was collected from a variety of online CPU benchmarking sources~\cite{spec,passmark} and press/product release information from prominent CPU designers. We consider CPUs with the same architecture design but different generations, performance tiers, or optimizations as different processors. For each processor, we show the maximum reported clock rate, including possible turbo clock rates. For processors containing multiple cores operating at different clock rates, we show the clock rate of the fastest core. We restrict our attention to general purpose CPUs designed for desktops, laptops, servers, tablets, smartphones, wearable devices, etc; however, we exclude application-specific hardware accelerators such as graphics or tensor processing units, which typically have far lower clock rates compared to CPUs. The maximum CPU clock rate used for the dashed vertical line in Fig.~\ref{fig:2}b is based on the current (as of 14 March 2024) CPU-Z overclocking world-record of 9117.75 MHz~\cite{hwbot}.  
\subsection*{Acknowledgments}
The authors acknowledge support from ARO grant no. W911NF-23-1-0048, NSF grant no. 1918549, Center for Sensing to Intelligence at Caltech, and NASA/JPL. The authors thank Martin M. Fejer and Carsten Langrock for providing the PPLN waveguide, and B. Volkan Gurses for lending RF amplifiers. G.H.Y.L acknowledges support from the Quad Fellowship. 
\subsection*{Author Contributions}
All authors contributed to this manuscript.
\subsection*{Data Availability}
The data used to generate the plots and results in this paper are available from the corresponding author upon reasonable request.
\subsection*{Code Availability}
The code used to analyze the data and generate the plots for this paper is available from
the corresponding author upon reasonable request.
\subsection*{Competing Interests}
AM has financial interest in PINC Technologies Inc., which is developing photonic integrated nonlinear circuits. The remaining authors declare no competing interests.
\bibliography{main_refs}

\begin{thebibliography}{68}%
\makeatletter
\providecommand \@ifxundefined [1]{%
 \@ifx{#1\undefined}
}%
\providecommand \@ifnum [1]{%
 \ifnum #1\expandafter \@firstoftwo
 \else \expandafter \@secondoftwo
 \fi
}%
\providecommand \@ifx [1]{%
 \ifx #1\expandafter \@firstoftwo
 \else \expandafter \@secondoftwo
 \fi
}%
\providecommand \natexlab [1]{#1}%
\providecommand \enquote  [1]{``#1''}%
\providecommand \bibnamefont  [1]{#1}%
\providecommand \bibfnamefont [1]{#1}%
\providecommand \citenamefont [1]{#1}%
\providecommand \href@noop [0]{\@secondoftwo}%
\providecommand \href [0]{\begingroup \@sanitize@url \@href}%
\providecommand \@href[1]{\@@startlink{#1}\@@href}%
\providecommand \@@href[1]{\endgroup#1\@@endlink}%
\providecommand \@sanitize@url [0]{\catcode `\\12\catcode `\$12\catcode `\&12\catcode `\#12\catcode `\^12\catcode `\_12\catcode `\%12\relax}%
\providecommand \@@startlink[1]{}%
\providecommand \@@endlink[0]{}%
\providecommand \url  [0]{\begingroup\@sanitize@url \@url }%
\providecommand \@url [1]{\endgroup\@href {#1}{\urlprefix }}%
\providecommand \urlprefix  [0]{URL }%
\providecommand \Eprint [0]{\href }%
\providecommand \doibase [0]{https://doi.org/}%
\providecommand \selectlanguage [0]{\@gobble}%
\providecommand \bibinfo  [0]{\@secondoftwo}%
\providecommand \bibfield  [0]{\@secondoftwo}%
\providecommand \translation [1]{[#1]}%
\providecommand \BibitemOpen [0]{}%
\providecommand \bibitemStop [0]{}%
\providecommand \bibitemNoStop [0]{.\EOS\space}%
\providecommand \EOS [0]{\spacefactor3000\relax}%
\providecommand \BibitemShut  [1]{\csname bibitem#1\endcsname}%
\let\auto@bib@innerbib\@empty
\bibitem [{\citenamefont {Hennessy}\ and\ \citenamefont {Patterson}(2011)}]{hennessy2011computer}%
  \BibitemOpen
  \bibfield  {author} {\bibinfo {author} {\bibfnamefont {J.~L.}\ \bibnamefont {Hennessy}}\ and\ \bibinfo {author} {\bibfnamefont {D.~A.}\ \bibnamefont {Patterson}},\ }\href@noop {} {\emph {\bibinfo {title} {Computer architecture: a quantitative approach}}}\ (\bibinfo  {publisher} {Elsevier},\ \bibinfo {year} {2011})\BibitemShut {NoStop}%
\bibitem [{\citenamefont {Bauer}(2009)}]{bauer2009origins}%
  \BibitemOpen
  \bibfield  {author} {\bibinfo {author} {\bibfnamefont {F.~L.}\ \bibnamefont {Bauer}},\ }\href@noop {} {\emph {\bibinfo {title} {Origins and foundations of computing: in cooperation with Heinz Nixdorf MuseumsForum}}}\ (\bibinfo  {publisher} {Springer Science \& Business Media},\ \bibinfo {year} {2009})\BibitemShut {NoStop}%
\bibitem [{\citenamefont {Burks}(1947)}]{burks1947electronic}%
  \BibitemOpen
  \bibfield  {author} {\bibinfo {author} {\bibfnamefont {A.~W.}\ \bibnamefont {Burks}},\ }\bibfield  {title} {\bibinfo {title} {Electronic computing circuits of the eniac},\ }\href@noop {} {\bibfield  {journal} {\bibinfo  {journal} {Proceedings of the IRE}\ }\textbf {\bibinfo {volume} {35}},\ \bibinfo {pages} {756} (\bibinfo {year} {1947})}\BibitemShut {NoStop}%
\bibitem [{\citenamefont {Schaller}(1997)}]{schaller1997moore}%
  \BibitemOpen
  \bibfield  {author} {\bibinfo {author} {\bibfnamefont {R.~R.}\ \bibnamefont {Schaller}},\ }\bibfield  {title} {\bibinfo {title} {Moore's law: past, present and future},\ }\href@noop {} {\bibfield  {journal} {\bibinfo  {journal} {IEEE Spectrum}\ }\textbf {\bibinfo {volume} {34}},\ \bibinfo {pages} {52} (\bibinfo {year} {1997})}\BibitemShut {NoStop}%
\bibitem [{\citenamefont {Dennard}\ \emph {et~al.}(1974)\citenamefont {Dennard}, \citenamefont {Gaensslen}, \citenamefont {Yu}, \citenamefont {Rideout}, \citenamefont {Bassous},\ and\ \citenamefont {LeBlanc}}]{dennard1974design}%
  \BibitemOpen
  \bibfield  {author} {\bibinfo {author} {\bibfnamefont {R.~H.}\ \bibnamefont {Dennard}}, \bibinfo {author} {\bibfnamefont {F.~H.}\ \bibnamefont {Gaensslen}}, \bibinfo {author} {\bibfnamefont {H.-N.}\ \bibnamefont {Yu}}, \bibinfo {author} {\bibfnamefont {V.~L.}\ \bibnamefont {Rideout}}, \bibinfo {author} {\bibfnamefont {E.}~\bibnamefont {Bassous}},\ and\ \bibinfo {author} {\bibfnamefont {A.~R.}\ \bibnamefont {LeBlanc}},\ }\bibfield  {title} {\bibinfo {title} {Design of ion-implanted mosfet's with very small physical dimensions},\ }\href@noop {} {\bibfield  {journal} {\bibinfo  {journal} {IEEE Journal of solid-state circuits}\ }\textbf {\bibinfo {volume} {9}},\ \bibinfo {pages} {256} (\bibinfo {year} {1974})}\BibitemShut {NoStop}%
\bibitem [{\citenamefont {Backus}(1978)}]{backus1978can}%
  \BibitemOpen
  \bibfield  {author} {\bibinfo {author} {\bibfnamefont {J.}~\bibnamefont {Backus}},\ }\bibfield  {title} {\bibinfo {title} {Can programming be liberated from the von neumann style? a functional style and its algebra of programs},\ }\href@noop {} {\bibfield  {journal} {\bibinfo  {journal} {Communications of the ACM}\ }\textbf {\bibinfo {volume} {21}},\ \bibinfo {pages} {613} (\bibinfo {year} {1978})}\BibitemShut {NoStop}%
\bibitem [{\citenamefont {McMahon}(2023)}]{mcmahon2023physics}%
  \BibitemOpen
  \bibfield  {author} {\bibinfo {author} {\bibfnamefont {P.~L.}\ \bibnamefont {McMahon}},\ }\bibfield  {title} {\bibinfo {title} {The physics of optical computing},\ }\href@noop {} {\bibfield  {journal} {\bibinfo  {journal} {Nature Reviews Physics}\ }\textbf {\bibinfo {volume} {5}},\ \bibinfo {pages} {717} (\bibinfo {year} {2023})}\BibitemShut {NoStop}%
\bibitem [{\citenamefont {Heinz}\ \emph {et~al.}(1970)\citenamefont {Heinz}, \citenamefont {Artman},\ and\ \citenamefont {Lee}}]{heinz1970matrix}%
  \BibitemOpen
  \bibfield  {author} {\bibinfo {author} {\bibfnamefont {R.}~\bibnamefont {Heinz}}, \bibinfo {author} {\bibfnamefont {J.}~\bibnamefont {Artman}},\ and\ \bibinfo {author} {\bibfnamefont {S.}~\bibnamefont {Lee}},\ }\bibfield  {title} {\bibinfo {title} {Matrix multiplication by optical methods},\ }\href@noop {} {\bibfield  {journal} {\bibinfo  {journal} {Applied Optics}\ }\textbf {\bibinfo {volume} {9}},\ \bibinfo {pages} {2161} (\bibinfo {year} {1970})}\BibitemShut {NoStop}%
\bibitem [{\citenamefont {Xu}\ \emph {et~al.}(2021)\citenamefont {Xu}, \citenamefont {Tan}, \citenamefont {Corcoran}, \citenamefont {Wu}, \citenamefont {Boes}, \citenamefont {Nguyen}, \citenamefont {Chu}, \citenamefont {Little}, \citenamefont {Hicks}, \citenamefont {Morandotti} \emph {et~al.}}]{xu202111}%
  \BibitemOpen
  \bibfield  {author} {\bibinfo {author} {\bibfnamefont {X.}~\bibnamefont {Xu}}, \bibinfo {author} {\bibfnamefont {M.}~\bibnamefont {Tan}}, \bibinfo {author} {\bibfnamefont {B.}~\bibnamefont {Corcoran}}, \bibinfo {author} {\bibfnamefont {J.}~\bibnamefont {Wu}}, \bibinfo {author} {\bibfnamefont {A.}~\bibnamefont {Boes}}, \bibinfo {author} {\bibfnamefont {T.~G.}\ \bibnamefont {Nguyen}}, \bibinfo {author} {\bibfnamefont {S.~T.}\ \bibnamefont {Chu}}, \bibinfo {author} {\bibfnamefont {B.~E.}\ \bibnamefont {Little}}, \bibinfo {author} {\bibfnamefont {D.~G.}\ \bibnamefont {Hicks}}, \bibinfo {author} {\bibfnamefont {R.}~\bibnamefont {Morandotti}}, \emph {et~al.},\ }\bibfield  {title} {\bibinfo {title} {11 tops photonic convolutional accelerator for optical neural networks},\ }\href@noop {} {\bibfield  {journal} {\bibinfo  {journal} {Nature}\ }\textbf {\bibinfo {volume} {589}},\ \bibinfo {pages} {44} (\bibinfo {year} {2021})}\BibitemShut {NoStop}%
\bibitem [{\citenamefont {Feldmann}\ \emph {et~al.}(2021)\citenamefont {Feldmann}, \citenamefont {Youngblood}, \citenamefont {Karpov}, \citenamefont {Gehring}, \citenamefont {Li}, \citenamefont {Stappers}, \citenamefont {Le~Gallo}, \citenamefont {Fu}, \citenamefont {Lukashchuk}, \citenamefont {Raja} \emph {et~al.}}]{feldmann2021parallel}%
  \BibitemOpen
  \bibfield  {author} {\bibinfo {author} {\bibfnamefont {J.}~\bibnamefont {Feldmann}}, \bibinfo {author} {\bibfnamefont {N.}~\bibnamefont {Youngblood}}, \bibinfo {author} {\bibfnamefont {M.}~\bibnamefont {Karpov}}, \bibinfo {author} {\bibfnamefont {H.}~\bibnamefont {Gehring}}, \bibinfo {author} {\bibfnamefont {X.}~\bibnamefont {Li}}, \bibinfo {author} {\bibfnamefont {M.}~\bibnamefont {Stappers}}, \bibinfo {author} {\bibfnamefont {M.}~\bibnamefont {Le~Gallo}}, \bibinfo {author} {\bibfnamefont {X.}~\bibnamefont {Fu}}, \bibinfo {author} {\bibfnamefont {A.}~\bibnamefont {Lukashchuk}}, \bibinfo {author} {\bibfnamefont {A.~S.}\ \bibnamefont {Raja}}, \emph {et~al.},\ }\bibfield  {title} {\bibinfo {title} {Parallel convolutional processing using an integrated photonic tensor core},\ }\href@noop {} {\bibfield  {journal} {\bibinfo  {journal} {Nature}\ }\textbf {\bibinfo {volume} {589}},\ \bibinfo {pages} {52} (\bibinfo {year} {2021})}\BibitemShut {NoStop}%
\bibitem [{\citenamefont {Li}\ \emph {et~al.}(2023)\citenamefont {Li}, \citenamefont {Sekine}, \citenamefont {Nehra}, \citenamefont {Gray}, \citenamefont {Ledezma}, \citenamefont {Guo},\ and\ \citenamefont {Marandi}}]{li2023all}%
  \BibitemOpen
  \bibfield  {author} {\bibinfo {author} {\bibfnamefont {G.~H.}\ \bibnamefont {Li}}, \bibinfo {author} {\bibfnamefont {R.}~\bibnamefont {Sekine}}, \bibinfo {author} {\bibfnamefont {R.}~\bibnamefont {Nehra}}, \bibinfo {author} {\bibfnamefont {R.~M.}\ \bibnamefont {Gray}}, \bibinfo {author} {\bibfnamefont {L.}~\bibnamefont {Ledezma}}, \bibinfo {author} {\bibfnamefont {Q.}~\bibnamefont {Guo}},\ and\ \bibinfo {author} {\bibfnamefont {A.}~\bibnamefont {Marandi}},\ }\bibfield  {title} {\bibinfo {title} {All-optical ultrafast relu function for energy-efficient nanophotonic deep learning},\ }\href@noop {} {\bibfield  {journal} {\bibinfo  {journal} {Nanophotonics}\ }\textbf {\bibinfo {volume} {12}},\ \bibinfo {pages} {847} (\bibinfo {year} {2023})}\BibitemShut {NoStop}%
\bibitem [{\citenamefont {Guo}\ \emph {et~al.}(2022)\citenamefont {Guo}, \citenamefont {Sekine}, \citenamefont {Ledezma}, \citenamefont {Nehra}, \citenamefont {Dean}, \citenamefont {Roy}, \citenamefont {Gray}, \citenamefont {Jahani},\ and\ \citenamefont {Marandi}}]{guo2022femtojoule}%
  \BibitemOpen
  \bibfield  {author} {\bibinfo {author} {\bibfnamefont {Q.}~\bibnamefont {Guo}}, \bibinfo {author} {\bibfnamefont {R.}~\bibnamefont {Sekine}}, \bibinfo {author} {\bibfnamefont {L.}~\bibnamefont {Ledezma}}, \bibinfo {author} {\bibfnamefont {R.}~\bibnamefont {Nehra}}, \bibinfo {author} {\bibfnamefont {D.~J.}\ \bibnamefont {Dean}}, \bibinfo {author} {\bibfnamefont {A.}~\bibnamefont {Roy}}, \bibinfo {author} {\bibfnamefont {R.~M.}\ \bibnamefont {Gray}}, \bibinfo {author} {\bibfnamefont {S.}~\bibnamefont {Jahani}},\ and\ \bibinfo {author} {\bibfnamefont {A.}~\bibnamefont {Marandi}},\ }\bibfield  {title} {\bibinfo {title} {Femtojoule femtosecond all-optical switching in lithium niobate nanophotonics},\ }\href@noop {} {\bibfield  {journal} {\bibinfo  {journal} {Nature Photonics}\ }\textbf {\bibinfo {volume} {16}},\ \bibinfo {pages} {625} (\bibinfo {year} {2022})}\BibitemShut {NoStop}%
\bibitem [{\citenamefont {Mourgias-Alexandris}\ \emph {et~al.}(2019)\citenamefont {Mourgias-Alexandris}, \citenamefont {Tsakyridis}, \citenamefont {Passalis}, \citenamefont {Tefas}, \citenamefont {Vyrsokinos},\ and\ \citenamefont {Pleros}}]{mourgias2019all}%
  \BibitemOpen
  \bibfield  {author} {\bibinfo {author} {\bibfnamefont {G.}~\bibnamefont {Mourgias-Alexandris}}, \bibinfo {author} {\bibfnamefont {A.}~\bibnamefont {Tsakyridis}}, \bibinfo {author} {\bibfnamefont {N.}~\bibnamefont {Passalis}}, \bibinfo {author} {\bibfnamefont {A.}~\bibnamefont {Tefas}}, \bibinfo {author} {\bibfnamefont {K.}~\bibnamefont {Vyrsokinos}},\ and\ \bibinfo {author} {\bibfnamefont {N.}~\bibnamefont {Pleros}},\ }\bibfield  {title} {\bibinfo {title} {An all-optical neuron with sigmoid activation function},\ }\href@noop {} {\bibfield  {journal} {\bibinfo  {journal} {Optics Express}\ }\textbf {\bibinfo {volume} {27}},\ \bibinfo {pages} {9620} (\bibinfo {year} {2019})}\BibitemShut {NoStop}%
\bibitem [{\citenamefont {Shen}\ \emph {et~al.}(2017)\citenamefont {Shen}, \citenamefont {Harris}, \citenamefont {Skirlo}, \citenamefont {Prabhu}, \citenamefont {Baehr-Jones}, \citenamefont {Hochberg}, \citenamefont {Sun}, \citenamefont {Zhao}, \citenamefont {Larochelle}, \citenamefont {Englund} \emph {et~al.}}]{shen2017deep}%
  \BibitemOpen
  \bibfield  {author} {\bibinfo {author} {\bibfnamefont {Y.}~\bibnamefont {Shen}}, \bibinfo {author} {\bibfnamefont {N.~C.}\ \bibnamefont {Harris}}, \bibinfo {author} {\bibfnamefont {S.}~\bibnamefont {Skirlo}}, \bibinfo {author} {\bibfnamefont {M.}~\bibnamefont {Prabhu}}, \bibinfo {author} {\bibfnamefont {T.}~\bibnamefont {Baehr-Jones}}, \bibinfo {author} {\bibfnamefont {M.}~\bibnamefont {Hochberg}}, \bibinfo {author} {\bibfnamefont {X.}~\bibnamefont {Sun}}, \bibinfo {author} {\bibfnamefont {S.}~\bibnamefont {Zhao}}, \bibinfo {author} {\bibfnamefont {H.}~\bibnamefont {Larochelle}}, \bibinfo {author} {\bibfnamefont {D.}~\bibnamefont {Englund}}, \emph {et~al.},\ }\bibfield  {title} {\bibinfo {title} {Deep learning with coherent nanophotonic circuits},\ }\href@noop {} {\bibfield  {journal} {\bibinfo  {journal} {Nature Photonics}\ }\textbf {\bibinfo {volume} {11}},\ \bibinfo {pages} {441} (\bibinfo {year} {2017})}\BibitemShut {NoStop}%
\bibitem [{\citenamefont {Ashtiani}\ \emph {et~al.}(2022)\citenamefont {Ashtiani}, \citenamefont {Geers},\ and\ \citenamefont {Aflatouni}}]{ashtiani2022chip}%
  \BibitemOpen
  \bibfield  {author} {\bibinfo {author} {\bibfnamefont {F.}~\bibnamefont {Ashtiani}}, \bibinfo {author} {\bibfnamefont {A.~J.}\ \bibnamefont {Geers}},\ and\ \bibinfo {author} {\bibfnamefont {F.}~\bibnamefont {Aflatouni}},\ }\bibfield  {title} {\bibinfo {title} {An on-chip photonic deep neural network for image classification},\ }\href@noop {} {\bibfield  {journal} {\bibinfo  {journal} {Nature}\ }\textbf {\bibinfo {volume} {606}},\ \bibinfo {pages} {501} (\bibinfo {year} {2022})}\BibitemShut {NoStop}%
\bibitem [{\citenamefont {Bandyopadhyay}\ \emph {et~al.}(2024)\citenamefont {Bandyopadhyay}, \citenamefont {Sludds}, \citenamefont {Krastanov}, \citenamefont {Hamerly}, \citenamefont {Harris}, \citenamefont {Bunandar}, \citenamefont {Streshinsky}, \citenamefont {Hochberg},\ and\ \citenamefont {Englund}}]{bandyopadhyay2024single}%
  \BibitemOpen
  \bibfield  {author} {\bibinfo {author} {\bibfnamefont {S.}~\bibnamefont {Bandyopadhyay}}, \bibinfo {author} {\bibfnamefont {A.}~\bibnamefont {Sludds}}, \bibinfo {author} {\bibfnamefont {S.}~\bibnamefont {Krastanov}}, \bibinfo {author} {\bibfnamefont {R.}~\bibnamefont {Hamerly}}, \bibinfo {author} {\bibfnamefont {N.}~\bibnamefont {Harris}}, \bibinfo {author} {\bibfnamefont {D.}~\bibnamefont {Bunandar}}, \bibinfo {author} {\bibfnamefont {M.}~\bibnamefont {Streshinsky}}, \bibinfo {author} {\bibfnamefont {M.}~\bibnamefont {Hochberg}},\ and\ \bibinfo {author} {\bibfnamefont {D.}~\bibnamefont {Englund}},\ }\bibfield  {title} {\bibinfo {title} {Single-chip photonic deep neural network with forward-only training},\ }\href@noop {} {\bibfield  {journal} {\bibinfo  {journal} {Nature Photonics}\ ,\ \bibinfo {pages} {1}} (\bibinfo {year} {2024})}\BibitemShut {NoStop}%
\bibitem [{\citenamefont {Feldmann}\ \emph {et~al.}(2019)\citenamefont {Feldmann}, \citenamefont {Youngblood}, \citenamefont {Wright}, \citenamefont {Bhaskaran},\ and\ \citenamefont {Pernice}}]{feldmann2019all}%
  \BibitemOpen
  \bibfield  {author} {\bibinfo {author} {\bibfnamefont {J.}~\bibnamefont {Feldmann}}, \bibinfo {author} {\bibfnamefont {N.}~\bibnamefont {Youngblood}}, \bibinfo {author} {\bibfnamefont {C.~D.}\ \bibnamefont {Wright}}, \bibinfo {author} {\bibfnamefont {H.}~\bibnamefont {Bhaskaran}},\ and\ \bibinfo {author} {\bibfnamefont {W.~H.}\ \bibnamefont {Pernice}},\ }\bibfield  {title} {\bibinfo {title} {All-optical spiking neurosynaptic networks with self-learning capabilities},\ }\href@noop {} {\bibfield  {journal} {\bibinfo  {journal} {Nature}\ }\textbf {\bibinfo {volume} {569}},\ \bibinfo {pages} {208} (\bibinfo {year} {2019})}\BibitemShut {NoStop}%
\bibitem [{\citenamefont {Tait}\ \emph {et~al.}(2017)\citenamefont {Tait}, \citenamefont {De~Lima}, \citenamefont {Zhou}, \citenamefont {Wu}, \citenamefont {Nahmias}, \citenamefont {Shastri},\ and\ \citenamefont {Prucnal}}]{tait2017neuromorphic}%
  \BibitemOpen
  \bibfield  {author} {\bibinfo {author} {\bibfnamefont {A.~N.}\ \bibnamefont {Tait}}, \bibinfo {author} {\bibfnamefont {T.~F.}\ \bibnamefont {De~Lima}}, \bibinfo {author} {\bibfnamefont {E.}~\bibnamefont {Zhou}}, \bibinfo {author} {\bibfnamefont {A.~X.}\ \bibnamefont {Wu}}, \bibinfo {author} {\bibfnamefont {M.~A.}\ \bibnamefont {Nahmias}}, \bibinfo {author} {\bibfnamefont {B.~J.}\ \bibnamefont {Shastri}},\ and\ \bibinfo {author} {\bibfnamefont {P.~R.}\ \bibnamefont {Prucnal}},\ }\bibfield  {title} {\bibinfo {title} {Neuromorphic photonic networks using silicon photonic weight banks},\ }\href@noop {} {\bibfield  {journal} {\bibinfo  {journal} {Scientific Reports}\ }\textbf {\bibinfo {volume} {7}},\ \bibinfo {pages} {7430} (\bibinfo {year} {2017})}\BibitemShut {NoStop}%
\bibitem [{\citenamefont {Prucnal}\ and\ \citenamefont {Shastri}(2017)}]{prucnal2017neuromorphic}%
  \BibitemOpen
  \bibfield  {author} {\bibinfo {author} {\bibfnamefont {P.~R.}\ \bibnamefont {Prucnal}}\ and\ \bibinfo {author} {\bibfnamefont {B.~J.}\ \bibnamefont {Shastri}},\ }\href@noop {} {\emph {\bibinfo {title} {Neuromorphic photonics}}}\ (\bibinfo  {publisher} {CRC press},\ \bibinfo {year} {2017})\BibitemShut {NoStop}%
\bibitem [{\citenamefont {Brunner}\ \emph {et~al.}(2025)\citenamefont {Brunner}, \citenamefont {Shastri}, \citenamefont {Qadasi}, \citenamefont {Ballani}, \citenamefont {Barbay}, \citenamefont {Biasi}, \citenamefont {Bienstman}, \citenamefont {Bilodeau}, \citenamefont {Bogaerts}, \citenamefont {B{\"o}hm} \emph {et~al.}}]{brunner2025roadmap}%
  \BibitemOpen
  \bibfield  {author} {\bibinfo {author} {\bibfnamefont {D.}~\bibnamefont {Brunner}}, \bibinfo {author} {\bibfnamefont {B.~J.}\ \bibnamefont {Shastri}}, \bibinfo {author} {\bibfnamefont {M.~A.~A.}\ \bibnamefont {Qadasi}}, \bibinfo {author} {\bibfnamefont {H.}~\bibnamefont {Ballani}}, \bibinfo {author} {\bibfnamefont {S.}~\bibnamefont {Barbay}}, \bibinfo {author} {\bibfnamefont {S.}~\bibnamefont {Biasi}}, \bibinfo {author} {\bibfnamefont {P.}~\bibnamefont {Bienstman}}, \bibinfo {author} {\bibfnamefont {S.}~\bibnamefont {Bilodeau}}, \bibinfo {author} {\bibfnamefont {W.}~\bibnamefont {Bogaerts}}, \bibinfo {author} {\bibfnamefont {F.}~\bibnamefont {B{\"o}hm}}, \emph {et~al.},\ }\bibfield  {title} {\bibinfo {title} {Roadmap on neuromorphic photonics},\ }\href@noop {} {\bibfield  {journal} {\bibinfo  {journal} {arXiv preprint arXiv:2501.07917}\ } (\bibinfo {year} {2025})}\BibitemShut {NoStop}%
\bibitem [{\citenamefont {Marandi}\ \emph {et~al.}(2014)\citenamefont {Marandi}, \citenamefont {Wang}, \citenamefont {Takata}, \citenamefont {Byer},\ and\ \citenamefont {Yamamoto}}]{marandi2014network}%
  \BibitemOpen
  \bibfield  {author} {\bibinfo {author} {\bibfnamefont {A.}~\bibnamefont {Marandi}}, \bibinfo {author} {\bibfnamefont {Z.}~\bibnamefont {Wang}}, \bibinfo {author} {\bibfnamefont {K.}~\bibnamefont {Takata}}, \bibinfo {author} {\bibfnamefont {R.~L.}\ \bibnamefont {Byer}},\ and\ \bibinfo {author} {\bibfnamefont {Y.}~\bibnamefont {Yamamoto}},\ }\bibfield  {title} {\bibinfo {title} {Network of time-multiplexed optical parametric oscillators as a coherent ising machine},\ }\href@noop {} {\bibfield  {journal} {\bibinfo  {journal} {Nature Photonics}\ }\textbf {\bibinfo {volume} {8}},\ \bibinfo {pages} {937} (\bibinfo {year} {2014})}\BibitemShut {NoStop}%
\bibitem [{\citenamefont {Inagaki}\ \emph {et~al.}(2016)\citenamefont {Inagaki}, \citenamefont {Haribara}, \citenamefont {Igarashi}, \citenamefont {Sonobe}, \citenamefont {Tamate}, \citenamefont {Honjo}, \citenamefont {Marandi}, \citenamefont {McMahon}, \citenamefont {Umeki}, \citenamefont {Enbutsu} \emph {et~al.}}]{inagaki2016coherent}%
  \BibitemOpen
  \bibfield  {author} {\bibinfo {author} {\bibfnamefont {T.}~\bibnamefont {Inagaki}}, \bibinfo {author} {\bibfnamefont {Y.}~\bibnamefont {Haribara}}, \bibinfo {author} {\bibfnamefont {K.}~\bibnamefont {Igarashi}}, \bibinfo {author} {\bibfnamefont {T.}~\bibnamefont {Sonobe}}, \bibinfo {author} {\bibfnamefont {S.}~\bibnamefont {Tamate}}, \bibinfo {author} {\bibfnamefont {T.}~\bibnamefont {Honjo}}, \bibinfo {author} {\bibfnamefont {A.}~\bibnamefont {Marandi}}, \bibinfo {author} {\bibfnamefont {P.~L.}\ \bibnamefont {McMahon}}, \bibinfo {author} {\bibfnamefont {T.}~\bibnamefont {Umeki}}, \bibinfo {author} {\bibfnamefont {K.}~\bibnamefont {Enbutsu}}, \emph {et~al.},\ }\bibfield  {title} {\bibinfo {title} {A coherent ising machine for 2000-node optimization problems},\ }\href@noop {} {\bibfield  {journal} {\bibinfo  {journal} {Science}\ }\textbf {\bibinfo {volume} {354}},\ \bibinfo {pages} {603} (\bibinfo {year} {2016})}\BibitemShut {NoStop}%
\bibitem [{\citenamefont {Honjo}\ \emph {et~al.}(2021)\citenamefont {Honjo}, \citenamefont {Sonobe}, \citenamefont {Inaba}, \citenamefont {Inagaki}, \citenamefont {Ikuta}, \citenamefont {Yamada}, \citenamefont {Kazama}, \citenamefont {Enbutsu}, \citenamefont {Umeki}, \citenamefont {Kasahara} \emph {et~al.}}]{honjo2021100}%
  \BibitemOpen
  \bibfield  {author} {\bibinfo {author} {\bibfnamefont {T.}~\bibnamefont {Honjo}}, \bibinfo {author} {\bibfnamefont {T.}~\bibnamefont {Sonobe}}, \bibinfo {author} {\bibfnamefont {K.}~\bibnamefont {Inaba}}, \bibinfo {author} {\bibfnamefont {T.}~\bibnamefont {Inagaki}}, \bibinfo {author} {\bibfnamefont {T.}~\bibnamefont {Ikuta}}, \bibinfo {author} {\bibfnamefont {Y.}~\bibnamefont {Yamada}}, \bibinfo {author} {\bibfnamefont {T.}~\bibnamefont {Kazama}}, \bibinfo {author} {\bibfnamefont {K.}~\bibnamefont {Enbutsu}}, \bibinfo {author} {\bibfnamefont {T.}~\bibnamefont {Umeki}}, \bibinfo {author} {\bibfnamefont {R.}~\bibnamefont {Kasahara}}, \emph {et~al.},\ }\bibfield  {title} {\bibinfo {title} {100,000-spin coherent ising machine},\ }\href@noop {} {\bibfield  {journal} {\bibinfo  {journal} {Science advances}\ }\textbf {\bibinfo {volume} {7}},\ \bibinfo {pages} {eabh0952} (\bibinfo {year} {2021})}\BibitemShut {NoStop}%
\bibitem [{\citenamefont {Parto}\ \emph {et~al.}(2020)\citenamefont {Parto}, \citenamefont {Hayenga}, \citenamefont {Marandi}, \citenamefont {Christodoulides},\ and\ \citenamefont {Khajavikhan}}]{parto2020realizing}%
  \BibitemOpen
  \bibfield  {author} {\bibinfo {author} {\bibfnamefont {M.}~\bibnamefont {Parto}}, \bibinfo {author} {\bibfnamefont {W.}~\bibnamefont {Hayenga}}, \bibinfo {author} {\bibfnamefont {A.}~\bibnamefont {Marandi}}, \bibinfo {author} {\bibfnamefont {D.~N.}\ \bibnamefont {Christodoulides}},\ and\ \bibinfo {author} {\bibfnamefont {M.}~\bibnamefont {Khajavikhan}},\ }\bibfield  {title} {\bibinfo {title} {Realizing spin hamiltonians in nanoscale active photonic lattices},\ }\href@noop {} {\bibfield  {journal} {\bibinfo  {journal} {Nature Materials}\ }\textbf {\bibinfo {volume} {19}},\ \bibinfo {pages} {725} (\bibinfo {year} {2020})}\BibitemShut {NoStop}%
\bibitem [{\citenamefont {Lin}\ \emph {et~al.}(2018)\citenamefont {Lin}, \citenamefont {Rivenson}, \citenamefont {Yardimci}, \citenamefont {Veli}, \citenamefont {Luo}, \citenamefont {Jarrahi},\ and\ \citenamefont {Ozcan}}]{lin2018all}%
  \BibitemOpen
  \bibfield  {author} {\bibinfo {author} {\bibfnamefont {X.}~\bibnamefont {Lin}}, \bibinfo {author} {\bibfnamefont {Y.}~\bibnamefont {Rivenson}}, \bibinfo {author} {\bibfnamefont {N.~T.}\ \bibnamefont {Yardimci}}, \bibinfo {author} {\bibfnamefont {M.}~\bibnamefont {Veli}}, \bibinfo {author} {\bibfnamefont {Y.}~\bibnamefont {Luo}}, \bibinfo {author} {\bibfnamefont {M.}~\bibnamefont {Jarrahi}},\ and\ \bibinfo {author} {\bibfnamefont {A.}~\bibnamefont {Ozcan}},\ }\bibfield  {title} {\bibinfo {title} {All-optical machine learning using diffractive deep neural networks},\ }\href@noop {} {\bibfield  {journal} {\bibinfo  {journal} {Science}\ }\textbf {\bibinfo {volume} {361}},\ \bibinfo {pages} {1004} (\bibinfo {year} {2018})}\BibitemShut {NoStop}%
\bibitem [{\citenamefont {Duport}\ \emph {et~al.}(2012)\citenamefont {Duport}, \citenamefont {Schneider}, \citenamefont {Smerieri}, \citenamefont {Haelterman},\ and\ \citenamefont {Massar}}]{duport2012all}%
  \BibitemOpen
  \bibfield  {author} {\bibinfo {author} {\bibfnamefont {F.}~\bibnamefont {Duport}}, \bibinfo {author} {\bibfnamefont {B.}~\bibnamefont {Schneider}}, \bibinfo {author} {\bibfnamefont {A.}~\bibnamefont {Smerieri}}, \bibinfo {author} {\bibfnamefont {M.}~\bibnamefont {Haelterman}},\ and\ \bibinfo {author} {\bibfnamefont {S.}~\bibnamefont {Massar}},\ }\bibfield  {title} {\bibinfo {title} {All-optical reservoir computing},\ }\href@noop {} {\bibfield  {journal} {\bibinfo  {journal} {Optics Express}\ }\textbf {\bibinfo {volume} {20}},\ \bibinfo {pages} {22783} (\bibinfo {year} {2012})}\BibitemShut {NoStop}%
\bibitem [{\citenamefont {Dejonckheere}\ \emph {et~al.}(2014)\citenamefont {Dejonckheere}, \citenamefont {Duport}, \citenamefont {Smerieri}, \citenamefont {Fang}, \citenamefont {Oudar}, \citenamefont {Haelterman},\ and\ \citenamefont {Massar}}]{dejonckheere2014all}%
  \BibitemOpen
  \bibfield  {author} {\bibinfo {author} {\bibfnamefont {A.}~\bibnamefont {Dejonckheere}}, \bibinfo {author} {\bibfnamefont {F.}~\bibnamefont {Duport}}, \bibinfo {author} {\bibfnamefont {A.}~\bibnamefont {Smerieri}}, \bibinfo {author} {\bibfnamefont {L.}~\bibnamefont {Fang}}, \bibinfo {author} {\bibfnamefont {J.-L.}\ \bibnamefont {Oudar}}, \bibinfo {author} {\bibfnamefont {M.}~\bibnamefont {Haelterman}},\ and\ \bibinfo {author} {\bibfnamefont {S.}~\bibnamefont {Massar}},\ }\bibfield  {title} {\bibinfo {title} {All-optical reservoir computer based on saturation of absorption},\ }\href@noop {} {\bibfield  {journal} {\bibinfo  {journal} {Optics express}\ }\textbf {\bibinfo {volume} {22}},\ \bibinfo {pages} {10868} (\bibinfo {year} {2014})}\BibitemShut {NoStop}%
\bibitem [{\citenamefont {Zuo}\ \emph {et~al.}(2019)\citenamefont {Zuo}, \citenamefont {Li}, \citenamefont {Zhao}, \citenamefont {Jiang}, \citenamefont {Chen}, \citenamefont {Chen}, \citenamefont {Jo}, \citenamefont {Liu},\ and\ \citenamefont {Du}}]{zuo2019all}%
  \BibitemOpen
  \bibfield  {author} {\bibinfo {author} {\bibfnamefont {Y.}~\bibnamefont {Zuo}}, \bibinfo {author} {\bibfnamefont {B.}~\bibnamefont {Li}}, \bibinfo {author} {\bibfnamefont {Y.}~\bibnamefont {Zhao}}, \bibinfo {author} {\bibfnamefont {Y.}~\bibnamefont {Jiang}}, \bibinfo {author} {\bibfnamefont {Y.-C.}\ \bibnamefont {Chen}}, \bibinfo {author} {\bibfnamefont {P.}~\bibnamefont {Chen}}, \bibinfo {author} {\bibfnamefont {G.-B.}\ \bibnamefont {Jo}}, \bibinfo {author} {\bibfnamefont {J.}~\bibnamefont {Liu}},\ and\ \bibinfo {author} {\bibfnamefont {S.}~\bibnamefont {Du}},\ }\bibfield  {title} {\bibinfo {title} {All-optical neural network with nonlinear activation functions},\ }\href@noop {} {\bibfield  {journal} {\bibinfo  {journal} {Optica}\ }\textbf {\bibinfo {volume} {6}},\ \bibinfo {pages} {1132} (\bibinfo {year} {2019})}\BibitemShut {NoStop}%
\bibitem [{\citenamefont {Medsker}\ and\ \citenamefont {Jain}(1999)}]{medsker1999recurrent}%
  \BibitemOpen
  \bibfield  {author} {\bibinfo {author} {\bibfnamefont {L.}~\bibnamefont {Medsker}}\ and\ \bibinfo {author} {\bibfnamefont {L.~C.}\ \bibnamefont {Jain}},\ }\href@noop {} {\emph {\bibinfo {title} {Recurrent neural networks: design and applications}}}\ (\bibinfo  {publisher} {CRC press},\ \bibinfo {year} {1999})\BibitemShut {NoStop}%
\bibitem [{\citenamefont {Siegelmann}\ and\ \citenamefont {Sontag}(1992)}]{siegelmann1992computational}%
  \BibitemOpen
  \bibfield  {author} {\bibinfo {author} {\bibfnamefont {H.~T.}\ \bibnamefont {Siegelmann}}\ and\ \bibinfo {author} {\bibfnamefont {E.~D.}\ \bibnamefont {Sontag}},\ }\bibfield  {title} {\bibinfo {title} {On the computational power of neural nets},\ }in\ \href@noop {} {\emph {\bibinfo {booktitle} {Proceedings of the fifth annual workshop on Computational learning theory}}}\ (\bibinfo {year} {1992})\ pp.\ \bibinfo {pages} {440--449}\BibitemShut {NoStop}%
\bibitem [{\citenamefont {Leefmans}\ \emph {et~al.}(2022)\citenamefont {Leefmans}, \citenamefont {Dutt}, \citenamefont {Williams}, \citenamefont {Yuan}, \citenamefont {Parto}, \citenamefont {Nori}, \citenamefont {Fan},\ and\ \citenamefont {Marandi}}]{leefmans2022topological}%
  \BibitemOpen
  \bibfield  {author} {\bibinfo {author} {\bibfnamefont {C.}~\bibnamefont {Leefmans}}, \bibinfo {author} {\bibfnamefont {A.}~\bibnamefont {Dutt}}, \bibinfo {author} {\bibfnamefont {J.}~\bibnamefont {Williams}}, \bibinfo {author} {\bibfnamefont {L.}~\bibnamefont {Yuan}}, \bibinfo {author} {\bibfnamefont {M.}~\bibnamefont {Parto}}, \bibinfo {author} {\bibfnamefont {F.}~\bibnamefont {Nori}}, \bibinfo {author} {\bibfnamefont {S.}~\bibnamefont {Fan}},\ and\ \bibinfo {author} {\bibfnamefont {A.}~\bibnamefont {Marandi}},\ }\bibfield  {title} {\bibinfo {title} {Topological dissipation in a time-multiplexed photonic resonator network},\ }\href@noop {} {\bibfield  {journal} {\bibinfo  {journal} {Nature Physics}\ }\textbf {\bibinfo {volume} {18}},\ \bibinfo {pages} {442} (\bibinfo {year} {2022})}\BibitemShut {NoStop}%
\bibitem [{\citenamefont {Bai}\ \emph {et~al.}(2023)\citenamefont {Bai}, \citenamefont {Xu}, \citenamefont {Tan}, \citenamefont {Sun}, \citenamefont {Li}, \citenamefont {Wu}, \citenamefont {Morandotti}, \citenamefont {Mitchell}, \citenamefont {Xu},\ and\ \citenamefont {Moss}}]{bai2023photonic}%
  \BibitemOpen
  \bibfield  {author} {\bibinfo {author} {\bibfnamefont {Y.}~\bibnamefont {Bai}}, \bibinfo {author} {\bibfnamefont {X.}~\bibnamefont {Xu}}, \bibinfo {author} {\bibfnamefont {M.}~\bibnamefont {Tan}}, \bibinfo {author} {\bibfnamefont {Y.}~\bibnamefont {Sun}}, \bibinfo {author} {\bibfnamefont {Y.}~\bibnamefont {Li}}, \bibinfo {author} {\bibfnamefont {J.}~\bibnamefont {Wu}}, \bibinfo {author} {\bibfnamefont {R.}~\bibnamefont {Morandotti}}, \bibinfo {author} {\bibfnamefont {A.}~\bibnamefont {Mitchell}}, \bibinfo {author} {\bibfnamefont {K.}~\bibnamefont {Xu}},\ and\ \bibinfo {author} {\bibfnamefont {D.~J.}\ \bibnamefont {Moss}},\ }\bibfield  {title} {\bibinfo {title} {Photonic multiplexing techniques for neuromorphic computing},\ }\href@noop {} {\bibfield  {journal} {\bibinfo  {journal} {Nanophotonics}\ }\textbf {\bibinfo {volume} {12}},\ \bibinfo {pages} {795} (\bibinfo {year} {2023})}\BibitemShut {NoStop}%
\bibitem [{\citenamefont {Fortier}\ and\ \citenamefont {Baumann}(2019)}]{fortier201920}%
  \BibitemOpen
  \bibfield  {author} {\bibinfo {author} {\bibfnamefont {T.}~\bibnamefont {Fortier}}\ and\ \bibinfo {author} {\bibfnamefont {E.}~\bibnamefont {Baumann}},\ }\bibfield  {title} {\bibinfo {title} {20 years of developments in optical frequency comb technology and applications},\ }\href@noop {} {\bibfield  {journal} {\bibinfo  {journal} {Communications Physics}\ }\textbf {\bibinfo {volume} {2}},\ \bibinfo {pages} {153} (\bibinfo {year} {2019})}\BibitemShut {NoStop}%
\bibitem [{\citenamefont {Langrock}\ and\ \citenamefont {Fejer}(2007)}]{langrock2007fiber}%
  \BibitemOpen
  \bibfield  {author} {\bibinfo {author} {\bibfnamefont {C.}~\bibnamefont {Langrock}}\ and\ \bibinfo {author} {\bibfnamefont {M.}~\bibnamefont {Fejer}},\ }\bibfield  {title} {\bibinfo {title} {Fiber-feedback continuous-wave and synchronously-pumped singly-resonant ring optical parametric oscillators using reverse-proton-exchanged periodically-poled lithium niobate waveguides},\ }\href@noop {} {\bibfield  {journal} {\bibinfo  {journal} {Optics Letters}\ }\textbf {\bibinfo {volume} {32}},\ \bibinfo {pages} {2263} (\bibinfo {year} {2007})}\BibitemShut {NoStop}%
\bibitem [{\citenamefont {Li}\ \emph {et~al.}(2024)\citenamefont {Li}, \citenamefont {Leefmans}, \citenamefont {Williams}, \citenamefont {Gray}, \citenamefont {Parto},\ and\ \citenamefont {Marandi}}]{li2024deep}%
  \BibitemOpen
  \bibfield  {author} {\bibinfo {author} {\bibfnamefont {G.~H.}\ \bibnamefont {Li}}, \bibinfo {author} {\bibfnamefont {C.~R.}\ \bibnamefont {Leefmans}}, \bibinfo {author} {\bibfnamefont {J.}~\bibnamefont {Williams}}, \bibinfo {author} {\bibfnamefont {R.~M.}\ \bibnamefont {Gray}}, \bibinfo {author} {\bibfnamefont {M.}~\bibnamefont {Parto}},\ and\ \bibinfo {author} {\bibfnamefont {A.}~\bibnamefont {Marandi}},\ }\bibfield  {title} {\bibinfo {title} {Deep learning with photonic neural cellular automata},\ }\href@noop {} {\bibfield  {journal} {\bibinfo  {journal} {Light: Science \& Applications}\ }\textbf {\bibinfo {volume} {13}},\ \bibinfo {pages} {283} (\bibinfo {year} {2024})}\BibitemShut {NoStop}%
\bibitem [{\citenamefont {Yuan}\ \emph {et~al.}(2023)\citenamefont {Yuan}, \citenamefont {Gao}, \citenamefont {Yu}, \citenamefont {Wang}, \citenamefont {Jin}, \citenamefont {Ji}, \citenamefont {Feshali}, \citenamefont {Paniccia}, \citenamefont {Bowers},\ and\ \citenamefont {Vahala}}]{yuan2023soliton}%
  \BibitemOpen
  \bibfield  {author} {\bibinfo {author} {\bibfnamefont {Z.}~\bibnamefont {Yuan}}, \bibinfo {author} {\bibfnamefont {M.}~\bibnamefont {Gao}}, \bibinfo {author} {\bibfnamefont {Y.}~\bibnamefont {Yu}}, \bibinfo {author} {\bibfnamefont {H.}~\bibnamefont {Wang}}, \bibinfo {author} {\bibfnamefont {W.}~\bibnamefont {Jin}}, \bibinfo {author} {\bibfnamefont {Q.-X.}\ \bibnamefont {Ji}}, \bibinfo {author} {\bibfnamefont {A.}~\bibnamefont {Feshali}}, \bibinfo {author} {\bibfnamefont {M.}~\bibnamefont {Paniccia}}, \bibinfo {author} {\bibfnamefont {J.}~\bibnamefont {Bowers}},\ and\ \bibinfo {author} {\bibfnamefont {K.}~\bibnamefont {Vahala}},\ }\bibfield  {title} {\bibinfo {title} {Soliton pulse pairs at multiple colours in normal dispersion microresonators},\ }\href@noop {} {\bibfield  {journal} {\bibinfo  {journal} {Nature Photonics}\ }\textbf {\bibinfo {volume} {17}},\ \bibinfo {pages} {977} (\bibinfo {year} {2023})}\BibitemShut {NoStop}%
\bibitem [{\citenamefont {Kippenberg}\ \emph {et~al.}(2018)\citenamefont {Kippenberg}, \citenamefont {Gaeta}, \citenamefont {Lipson},\ and\ \citenamefont {Gorodetsky}}]{doi:10.1126/science.aan8083}%
  \BibitemOpen
  \bibfield  {author} {\bibinfo {author} {\bibfnamefont {T.~J.}\ \bibnamefont {Kippenberg}}, \bibinfo {author} {\bibfnamefont {A.~L.}\ \bibnamefont {Gaeta}}, \bibinfo {author} {\bibfnamefont {M.}~\bibnamefont {Lipson}},\ and\ \bibinfo {author} {\bibfnamefont {M.~L.}\ \bibnamefont {Gorodetsky}},\ }\bibfield  {title} {\bibinfo {title} {Dissipative kerr solitons in optical microresonators},\ }\href@noop {} {\bibfield  {journal} {\bibinfo  {journal} {Science}\ }\textbf {\bibinfo {volume} {361}},\ \bibinfo {pages} {eaan8083} (\bibinfo {year} {2018})}\BibitemShut {NoStop}%
\bibitem [{\citenamefont {Herink}\ \emph {et~al.}(2017)\citenamefont {Herink}, \citenamefont {Kurtz}, \citenamefont {Jalali}, \citenamefont {Solli},\ and\ \citenamefont {Ropers}}]{herink2017real}%
  \BibitemOpen
  \bibfield  {author} {\bibinfo {author} {\bibfnamefont {G.}~\bibnamefont {Herink}}, \bibinfo {author} {\bibfnamefont {F.}~\bibnamefont {Kurtz}}, \bibinfo {author} {\bibfnamefont {B.}~\bibnamefont {Jalali}}, \bibinfo {author} {\bibfnamefont {D.~R.}\ \bibnamefont {Solli}},\ and\ \bibinfo {author} {\bibfnamefont {C.}~\bibnamefont {Ropers}},\ }\bibfield  {title} {\bibinfo {title} {Real-time spectral interferometry probes the internal dynamics of femtosecond soliton molecules},\ }\href@noop {} {\bibfield  {journal} {\bibinfo  {journal} {Science}\ }\textbf {\bibinfo {volume} {356}},\ \bibinfo {pages} {50} (\bibinfo {year} {2017})}\BibitemShut {NoStop}%
\bibitem [{\citenamefont {Foster}\ \emph {et~al.}(2008)\citenamefont {Foster}, \citenamefont {Salem}, \citenamefont {Geraghty}, \citenamefont {Turner-Foster}, \citenamefont {Lipson},\ and\ \citenamefont {Gaeta}}]{foster2008silicon}%
  \BibitemOpen
  \bibfield  {author} {\bibinfo {author} {\bibfnamefont {M.~A.}\ \bibnamefont {Foster}}, \bibinfo {author} {\bibfnamefont {R.}~\bibnamefont {Salem}}, \bibinfo {author} {\bibfnamefont {D.~F.}\ \bibnamefont {Geraghty}}, \bibinfo {author} {\bibfnamefont {A.~C.}\ \bibnamefont {Turner-Foster}}, \bibinfo {author} {\bibfnamefont {M.}~\bibnamefont {Lipson}},\ and\ \bibinfo {author} {\bibfnamefont {A.~L.}\ \bibnamefont {Gaeta}},\ }\bibfield  {title} {\bibinfo {title} {Silicon-chip-based ultrafast optical oscilloscope},\ }\href@noop {} {\bibfield  {journal} {\bibinfo  {journal} {Nature}\ }\textbf {\bibinfo {volume} {456}},\ \bibinfo {pages} {81} (\bibinfo {year} {2008})}\BibitemShut {NoStop}%
\bibitem [{\citenamefont {Goda}\ and\ \citenamefont {Jalali}(2013)}]{goda2013dispersive}%
  \BibitemOpen
  \bibfield  {author} {\bibinfo {author} {\bibfnamefont {K.}~\bibnamefont {Goda}}\ and\ \bibinfo {author} {\bibfnamefont {B.}~\bibnamefont {Jalali}},\ }\bibfield  {title} {\bibinfo {title} {Dispersive fourier transformation for fast continuous single-shot measurements},\ }\href@noop {} {\bibfield  {journal} {\bibinfo  {journal} {Nature Photonics}\ }\textbf {\bibinfo {volume} {7}},\ \bibinfo {pages} {102} (\bibinfo {year} {2013})}\BibitemShut {NoStop}%
\bibitem [{\citenamefont {Kane}\ and\ \citenamefont {Trebino}(1993)}]{kane1993single}%
  \BibitemOpen
  \bibfield  {author} {\bibinfo {author} {\bibfnamefont {D.~J.}\ \bibnamefont {Kane}}\ and\ \bibinfo {author} {\bibfnamefont {R.}~\bibnamefont {Trebino}},\ }\bibfield  {title} {\bibinfo {title} {Single-shot measurement of the intensity and phase of an arbitrary ultrashort pulse by using frequency-resolved optical gating},\ }\href@noop {} {\bibfield  {journal} {\bibinfo  {journal} {Optics Letters}\ }\textbf {\bibinfo {volume} {18}},\ \bibinfo {pages} {823} (\bibinfo {year} {1993})}\BibitemShut {NoStop}%
\bibitem [{\citenamefont {Yi}\ \emph {et~al.}(2018)\citenamefont {Yi}, \citenamefont {Yang}, \citenamefont {Yang},\ and\ \citenamefont {Vahala}}]{yi2018imaging}%
  \BibitemOpen
  \bibfield  {author} {\bibinfo {author} {\bibfnamefont {X.}~\bibnamefont {Yi}}, \bibinfo {author} {\bibfnamefont {Q.-F.}\ \bibnamefont {Yang}}, \bibinfo {author} {\bibfnamefont {K.~Y.}\ \bibnamefont {Yang}},\ and\ \bibinfo {author} {\bibfnamefont {K.}~\bibnamefont {Vahala}},\ }\bibfield  {title} {\bibinfo {title} {Imaging soliton dynamics in optical microcavities},\ }\href@noop {} {\bibfield  {journal} {\bibinfo  {journal} {Nature communications}\ }\textbf {\bibinfo {volume} {9}},\ \bibinfo {pages} {3565} (\bibinfo {year} {2018})}\BibitemShut {NoStop}%
\bibitem [{\citenamefont {Gray}\ \emph {et~al.}(2024{\natexlab{a}})\citenamefont {Gray}, \citenamefont {Liu}, \citenamefont {Zhou}, \citenamefont {Roy}, \citenamefont {Ledezma},\ and\ \citenamefont {Marandi}}]{gray2024quadratic}%
  \BibitemOpen
  \bibfield  {author} {\bibinfo {author} {\bibfnamefont {R.~M.}\ \bibnamefont {Gray}}, \bibinfo {author} {\bibfnamefont {M.}~\bibnamefont {Liu}}, \bibinfo {author} {\bibfnamefont {S.}~\bibnamefont {Zhou}}, \bibinfo {author} {\bibfnamefont {A.}~\bibnamefont {Roy}}, \bibinfo {author} {\bibfnamefont {L.}~\bibnamefont {Ledezma}},\ and\ \bibinfo {author} {\bibfnamefont {A.}~\bibnamefont {Marandi}},\ }\bibfield  {title} {\bibinfo {title} {Quadratic-soliton-enhanced mid-ir molecular sensing},\ }\href@noop {} {\bibfield  {journal} {\bibinfo  {journal} {Nature Communications}\ }\textbf {\bibinfo {volume} {15}},\ \bibinfo {pages} {9086} (\bibinfo {year} {2024}{\natexlab{a}})}\BibitemShut {NoStop}%
\bibitem [{\citenamefont {Gomber}\ and\ \citenamefont {Haferkorn}(2013)}]{gomber2013high}%
  \BibitemOpen
  \bibfield  {author} {\bibinfo {author} {\bibfnamefont {P.}~\bibnamefont {Gomber}}\ and\ \bibinfo {author} {\bibfnamefont {M.}~\bibnamefont {Haferkorn}},\ }\bibfield  {title} {\bibinfo {title} {High-frequency-trading: High-frequency-trading technologies and their implications for electronic securities trading},\ }\href@noop {} {\bibfield  {journal} {\bibinfo  {journal} {Business \& Information Systems Engineering}\ }\textbf {\bibinfo {volume} {5}},\ \bibinfo {pages} {97} (\bibinfo {year} {2013})}\BibitemShut {NoStop}%
\bibitem [{\citenamefont {Gligorov}\ and\ \citenamefont {Williams}(2013)}]{gligorov2013efficient}%
  \BibitemOpen
  \bibfield  {author} {\bibinfo {author} {\bibfnamefont {V.~V.}\ \bibnamefont {Gligorov}}\ and\ \bibinfo {author} {\bibfnamefont {M.}~\bibnamefont {Williams}},\ }\bibfield  {title} {\bibinfo {title} {Efficient, reliable and fast high-level triggering using a bonsai boosted decision tree},\ }\href@noop {} {\bibfield  {journal} {\bibinfo  {journal} {Journal of Instrumentation}\ }\textbf {\bibinfo {volume} {8}}\bibinfo  {number} { (02)},\ \bibinfo {pages} {P02013}}\BibitemShut {NoStop}%
\bibitem [{\citenamefont {Willner}\ \emph {et~al.}(2013)\citenamefont {Willner}, \citenamefont {Khaleghi}, \citenamefont {Chitgarha},\ and\ \citenamefont {Yilmaz}}]{willner2013all}%
  \BibitemOpen
\bibfield  {number} {  }\bibfield  {author} {\bibinfo {author} {\bibfnamefont {A.~E.}\ \bibnamefont {Willner}}, \bibinfo {author} {\bibfnamefont {S.}~\bibnamefont {Khaleghi}}, \bibinfo {author} {\bibfnamefont {M.~R.}\ \bibnamefont {Chitgarha}},\ and\ \bibinfo {author} {\bibfnamefont {O.~F.}\ \bibnamefont {Yilmaz}},\ }\bibfield  {title} {\bibinfo {title} {All-optical signal processing},\ }\href@noop {} {\bibfield  {journal} {\bibinfo  {journal} {Journal of Lightwave Technology}\ }\textbf {\bibinfo {volume} {32}},\ \bibinfo {pages} {660} (\bibinfo {year} {2013})}\BibitemShut {NoStop}%
\bibitem [{\citenamefont {Choi}\ \emph {et~al.}(2024)\citenamefont {Choi}, \citenamefont {Salamin}, \citenamefont {Roques-Carmes}, \citenamefont {Dangovski}, \citenamefont {Luo}, \citenamefont {Chen}, \citenamefont {Horodynski}, \citenamefont {Sloan}, \citenamefont {Uddin},\ and\ \citenamefont {Solja{\v{c}}i{\'c}}}]{choi2024photonic}%
  \BibitemOpen
  \bibfield  {author} {\bibinfo {author} {\bibfnamefont {S.}~\bibnamefont {Choi}}, \bibinfo {author} {\bibfnamefont {Y.}~\bibnamefont {Salamin}}, \bibinfo {author} {\bibfnamefont {C.}~\bibnamefont {Roques-Carmes}}, \bibinfo {author} {\bibfnamefont {R.}~\bibnamefont {Dangovski}}, \bibinfo {author} {\bibfnamefont {D.}~\bibnamefont {Luo}}, \bibinfo {author} {\bibfnamefont {Z.}~\bibnamefont {Chen}}, \bibinfo {author} {\bibfnamefont {M.}~\bibnamefont {Horodynski}}, \bibinfo {author} {\bibfnamefont {J.}~\bibnamefont {Sloan}}, \bibinfo {author} {\bibfnamefont {S.~Z.}\ \bibnamefont {Uddin}},\ and\ \bibinfo {author} {\bibfnamefont {M.}~\bibnamefont {Solja{\v{c}}i{\'c}}},\ }\bibfield  {title} {\bibinfo {title} {Photonic probabilistic machine learning using quantum vacuum noise},\ }\href@noop {} {\bibfield  {journal} {\bibinfo  {journal} {Nature Communications}\ }\textbf {\bibinfo {volume} {15}},\ \bibinfo {pages} {7760} (\bibinfo {year} {2024})}\BibitemShut {NoStop}%
\bibitem [{\citenamefont {Roques-Carmes}\ \emph {et~al.}(2023)\citenamefont {Roques-Carmes}, \citenamefont {Salamin}, \citenamefont {Sloan}, \citenamefont {Choi}, \citenamefont {Velez}, \citenamefont {Koskas}, \citenamefont {Rivera}, \citenamefont {Kooi}, \citenamefont {Joannopoulos},\ and\ \citenamefont {Solja{\v{c}}i{\'c}}}]{roques2023biasing}%
  \BibitemOpen
  \bibfield  {author} {\bibinfo {author} {\bibfnamefont {C.}~\bibnamefont {Roques-Carmes}}, \bibinfo {author} {\bibfnamefont {Y.}~\bibnamefont {Salamin}}, \bibinfo {author} {\bibfnamefont {J.}~\bibnamefont {Sloan}}, \bibinfo {author} {\bibfnamefont {S.}~\bibnamefont {Choi}}, \bibinfo {author} {\bibfnamefont {G.}~\bibnamefont {Velez}}, \bibinfo {author} {\bibfnamefont {E.}~\bibnamefont {Koskas}}, \bibinfo {author} {\bibfnamefont {N.}~\bibnamefont {Rivera}}, \bibinfo {author} {\bibfnamefont {S.~E.}\ \bibnamefont {Kooi}}, \bibinfo {author} {\bibfnamefont {J.~D.}\ \bibnamefont {Joannopoulos}},\ and\ \bibinfo {author} {\bibfnamefont {M.}~\bibnamefont {Solja{\v{c}}i{\'c}}},\ }\bibfield  {title} {\bibinfo {title} {Biasing the quantum vacuum to control macroscopic probability distributions},\ }\href@noop {} {\bibfield  {journal} {\bibinfo  {journal} {Science}\ }\textbf {\bibinfo {volume} {381}},\ \bibinfo {pages} {205} (\bibinfo {year} {2023})}\BibitemShut {NoStop}%
\bibitem [{\citenamefont {Ho}\ \emph {et~al.}(2020)\citenamefont {Ho}, \citenamefont {Jain},\ and\ \citenamefont {Abbeel}}]{ho2020denoising}%
  \BibitemOpen
  \bibfield  {author} {\bibinfo {author} {\bibfnamefont {J.}~\bibnamefont {Ho}}, \bibinfo {author} {\bibfnamefont {A.}~\bibnamefont {Jain}},\ and\ \bibinfo {author} {\bibfnamefont {P.}~\bibnamefont {Abbeel}},\ }\bibfield  {title} {\bibinfo {title} {Denoising diffusion probabilistic models},\ }\href@noop {} {\bibfield  {journal} {\bibinfo  {journal} {Advances in neural information processing systems}\ }\textbf {\bibinfo {volume} {33}},\ \bibinfo {pages} {6840} (\bibinfo {year} {2020})}\BibitemShut {NoStop}%
\bibitem [{\citenamefont {Lipman}\ \emph {et~al.}(2022)\citenamefont {Lipman}, \citenamefont {Chen}, \citenamefont {Ben-Hamu}, \citenamefont {Nickel},\ and\ \citenamefont {Le}}]{lipman2022flow}%
  \BibitemOpen
  \bibfield  {author} {\bibinfo {author} {\bibfnamefont {Y.}~\bibnamefont {Lipman}}, \bibinfo {author} {\bibfnamefont {R.~T.}\ \bibnamefont {Chen}}, \bibinfo {author} {\bibfnamefont {H.}~\bibnamefont {Ben-Hamu}}, \bibinfo {author} {\bibfnamefont {M.}~\bibnamefont {Nickel}},\ and\ \bibinfo {author} {\bibfnamefont {M.}~\bibnamefont {Le}},\ }\bibfield  {title} {\bibinfo {title} {Flow matching for generative modeling},\ }\href@noop {} {\bibfield  {journal} {\bibinfo  {journal} {arXiv preprint arXiv:2210.02747}\ } (\bibinfo {year} {2022})}\BibitemShut {NoStop}%
\bibitem [{\citenamefont {Deng}(2012)}]{deng2012mnist}%
  \BibitemOpen
  \bibfield  {author} {\bibinfo {author} {\bibfnamefont {L.}~\bibnamefont {Deng}},\ }\bibfield  {title} {\bibinfo {title} {The mnist database of handwritten digit images for machine learning research [best of the web]},\ }\href@noop {} {\bibfield  {journal} {\bibinfo  {journal} {IEEE signal processing magazine}\ }\textbf {\bibinfo {volume} {29}},\ \bibinfo {pages} {141} (\bibinfo {year} {2012})}\BibitemShut {NoStop}%
\bibitem [{\citenamefont {Milonni}\ and\ \citenamefont {Eberly}(2010)}]{milonni2010laser}%
  \BibitemOpen
  \bibfield  {author} {\bibinfo {author} {\bibfnamefont {P.~W.}\ \bibnamefont {Milonni}}\ and\ \bibinfo {author} {\bibfnamefont {J.~H.}\ \bibnamefont {Eberly}},\ }\href@noop {} {\emph {\bibinfo {title} {Laser physics}}}\ (\bibinfo  {publisher} {John Wiley \& Sons},\ \bibinfo {year} {2010})\BibitemShut {NoStop}%
\bibitem [{\citenamefont {Wang}\ \emph {et~al.}(2018)\citenamefont {Wang}, \citenamefont {Zhang}, \citenamefont {Chen}, \citenamefont {Bertrand}, \citenamefont {Shams-Ansari}, \citenamefont {Chandrasekhar}, \citenamefont {Winzer},\ and\ \citenamefont {Lon{\v{c}}ar}}]{wang2018integrated}%
  \BibitemOpen
  \bibfield  {author} {\bibinfo {author} {\bibfnamefont {C.}~\bibnamefont {Wang}}, \bibinfo {author} {\bibfnamefont {M.}~\bibnamefont {Zhang}}, \bibinfo {author} {\bibfnamefont {X.}~\bibnamefont {Chen}}, \bibinfo {author} {\bibfnamefont {M.}~\bibnamefont {Bertrand}}, \bibinfo {author} {\bibfnamefont {A.}~\bibnamefont {Shams-Ansari}}, \bibinfo {author} {\bibfnamefont {S.}~\bibnamefont {Chandrasekhar}}, \bibinfo {author} {\bibfnamefont {P.}~\bibnamefont {Winzer}},\ and\ \bibinfo {author} {\bibfnamefont {M.}~\bibnamefont {Lon{\v{c}}ar}},\ }\bibfield  {title} {\bibinfo {title} {Integrated lithium niobate electro-optic modulators operating at cmos-compatible voltages},\ }\href@noop {} {\bibfield  {journal} {\bibinfo  {journal} {Nature}\ }\textbf {\bibinfo {volume} {562}},\ \bibinfo {pages} {101} (\bibinfo {year} {2018})}\BibitemShut {NoStop}%
\bibitem [{\citenamefont {Gray}\ \emph {et~al.}(2024{\natexlab{b}})\citenamefont {Gray}, \citenamefont {Zacharias}, \citenamefont {Chawlani}, \citenamefont {Ledezma}, \citenamefont {Sekine}, \citenamefont {Williams},\ and\ \citenamefont {Marandi}}]{gray2024soliton}%
  \BibitemOpen
  \bibfield  {author} {\bibinfo {author} {\bibfnamefont {R.~M.}\ \bibnamefont {Gray}}, \bibinfo {author} {\bibfnamefont {T.}~\bibnamefont {Zacharias}}, \bibinfo {author} {\bibfnamefont {R.}~\bibnamefont {Chawlani}}, \bibinfo {author} {\bibfnamefont {L.}~\bibnamefont {Ledezma}}, \bibinfo {author} {\bibfnamefont {R.}~\bibnamefont {Sekine}}, \bibinfo {author} {\bibfnamefont {J.~A.}\ \bibnamefont {Williams}},\ and\ \bibinfo {author} {\bibfnamefont {A.}~\bibnamefont {Marandi}},\ }\bibfield  {title} {\bibinfo {title} {Soliton pulse compression in lithium niobate nanophotonics},\ }in\ \href@noop {} {\emph {\bibinfo {booktitle} {2024 Conference on Lasers and Electro-Optics (CLEO)}}}\ (\bibinfo {organization} {IEEE},\ \bibinfo {year} {2024})\ pp.\ \bibinfo {pages} {1--2}\BibitemShut {NoStop}%
\bibitem [{\citenamefont {Roy}\ \emph {et~al.}(2023)\citenamefont {Roy}, \citenamefont {Ledezma}, \citenamefont {Costa}, \citenamefont {Gray}, \citenamefont {Sekine}, \citenamefont {Guo}, \citenamefont {Liu}, \citenamefont {Briggs},\ and\ \citenamefont {Marandi}}]{roy2023visible}%
  \BibitemOpen
  \bibfield  {author} {\bibinfo {author} {\bibfnamefont {A.}~\bibnamefont {Roy}}, \bibinfo {author} {\bibfnamefont {L.}~\bibnamefont {Ledezma}}, \bibinfo {author} {\bibfnamefont {L.}~\bibnamefont {Costa}}, \bibinfo {author} {\bibfnamefont {R.}~\bibnamefont {Gray}}, \bibinfo {author} {\bibfnamefont {R.}~\bibnamefont {Sekine}}, \bibinfo {author} {\bibfnamefont {Q.}~\bibnamefont {Guo}}, \bibinfo {author} {\bibfnamefont {M.}~\bibnamefont {Liu}}, \bibinfo {author} {\bibfnamefont {R.~M.}\ \bibnamefont {Briggs}},\ and\ \bibinfo {author} {\bibfnamefont {A.}~\bibnamefont {Marandi}},\ }\bibfield  {title} {\bibinfo {title} {Visible-to-mid-ir tunable frequency comb in nanophotonics},\ }\href@noop {} {\bibfield  {journal} {\bibinfo  {journal} {Nature Communications}\ }\textbf {\bibinfo {volume} {14}},\ \bibinfo {pages} {6549} (\bibinfo {year} {2023})}\BibitemShut {NoStop}%
\bibitem [{\citenamefont {Ledezma}\ \emph {et~al.}(2023)\citenamefont {Ledezma}, \citenamefont {Roy}, \citenamefont {Costa}, \citenamefont {Sekine}, \citenamefont {Gray}, \citenamefont {Guo}, \citenamefont {Nehra}, \citenamefont {Briggs},\ and\ \citenamefont {Marandi}}]{ledezma2023octave}%
  \BibitemOpen
  \bibfield  {author} {\bibinfo {author} {\bibfnamefont {L.}~\bibnamefont {Ledezma}}, \bibinfo {author} {\bibfnamefont {A.}~\bibnamefont {Roy}}, \bibinfo {author} {\bibfnamefont {L.}~\bibnamefont {Costa}}, \bibinfo {author} {\bibfnamefont {R.}~\bibnamefont {Sekine}}, \bibinfo {author} {\bibfnamefont {R.}~\bibnamefont {Gray}}, \bibinfo {author} {\bibfnamefont {Q.}~\bibnamefont {Guo}}, \bibinfo {author} {\bibfnamefont {R.}~\bibnamefont {Nehra}}, \bibinfo {author} {\bibfnamefont {R.~M.}\ \bibnamefont {Briggs}},\ and\ \bibinfo {author} {\bibfnamefont {A.}~\bibnamefont {Marandi}},\ }\bibfield  {title} {\bibinfo {title} {Octave-spanning tunable infrared parametric oscillators in nanophotonics},\ }\href@noop {} {\bibfield  {journal} {\bibinfo  {journal} {Science Advances}\ }\textbf {\bibinfo {volume} {9}},\ \bibinfo {pages} {eadf9711} (\bibinfo {year} {2023})}\BibitemShut {NoStop}%
\bibitem [{\citenamefont {Gray}\ \emph {et~al.}(2024{\natexlab{c}})\citenamefont {Gray}, \citenamefont {Sekine}, \citenamefont {Ledezma}, \citenamefont {Li}, \citenamefont {Zhou}, \citenamefont {Roy}, \citenamefont {Parto},\ and\ \citenamefont {Marandi}}]{gray2024large}%
  \BibitemOpen
  \bibfield  {author} {\bibinfo {author} {\bibfnamefont {R.~M.}\ \bibnamefont {Gray}}, \bibinfo {author} {\bibfnamefont {R.}~\bibnamefont {Sekine}}, \bibinfo {author} {\bibfnamefont {L.}~\bibnamefont {Ledezma}}, \bibinfo {author} {\bibfnamefont {G.~H.}\ \bibnamefont {Li}}, \bibinfo {author} {\bibfnamefont {S.}~\bibnamefont {Zhou}}, \bibinfo {author} {\bibfnamefont {A.}~\bibnamefont {Roy}}, \bibinfo {author} {\bibfnamefont {M.}~\bibnamefont {Parto}},\ and\ \bibinfo {author} {\bibfnamefont {A.}~\bibnamefont {Marandi}},\ }\bibfield  {title} {\bibinfo {title} {Large-scale time-multiplexed nanophotonic parametric oscillators},\ }\href@noop {} {\bibfield  {journal} {\bibinfo  {journal} {arXiv preprint arXiv:2405.17355}\ } (\bibinfo {year} {2024}{\natexlab{c}})}\BibitemShut {NoStop}%
\bibitem [{\citenamefont {Zhou}\ \emph {et~al.}(2023)\citenamefont {Zhou}, \citenamefont {Wu}, \citenamefont {Zhang}, \citenamefont {Yu},\ and\ \citenamefont {Fang}}]{zhou2023ultrafast}%
  \BibitemOpen
  \bibfield  {author} {\bibinfo {author} {\bibfnamefont {T.}~\bibnamefont {Zhou}}, \bibinfo {author} {\bibfnamefont {W.}~\bibnamefont {Wu}}, \bibinfo {author} {\bibfnamefont {J.}~\bibnamefont {Zhang}}, \bibinfo {author} {\bibfnamefont {S.}~\bibnamefont {Yu}},\ and\ \bibinfo {author} {\bibfnamefont {L.}~\bibnamefont {Fang}},\ }\bibfield  {title} {\bibinfo {title} {Ultrafast dynamic machine vision with spatiotemporal photonic computing},\ }\href@noop {} {\bibfield  {journal} {\bibinfo  {journal} {Science Advances}\ }\textbf {\bibinfo {volume} {9}},\ \bibinfo {pages} {eadg4391} (\bibinfo {year} {2023})}\BibitemShut {NoStop}%
\bibitem [{\citenamefont {Wang}\ \emph {et~al.}(2020)\citenamefont {Wang}, \citenamefont {Liang},\ and\ \citenamefont {Wang}}]{wang2020single}%
  \BibitemOpen
  \bibfield  {author} {\bibinfo {author} {\bibfnamefont {P.}~\bibnamefont {Wang}}, \bibinfo {author} {\bibfnamefont {J.}~\bibnamefont {Liang}},\ and\ \bibinfo {author} {\bibfnamefont {L.~V.}\ \bibnamefont {Wang}},\ }\bibfield  {title} {\bibinfo {title} {Single-shot ultrafast imaging attaining 70 trillion frames per second},\ }\href@noop {} {\bibfield  {journal} {\bibinfo  {journal} {Nature Communications}\ }\textbf {\bibinfo {volume} {11}},\ \bibinfo {pages} {2091} (\bibinfo {year} {2020})}\BibitemShut {NoStop}%
\bibitem [{\citenamefont {Maiuri}\ \emph {et~al.}(2019)\citenamefont {Maiuri}, \citenamefont {Garavelli},\ and\ \citenamefont {Cerullo}}]{maiuri2019ultrafast}%
  \BibitemOpen
  \bibfield  {author} {\bibinfo {author} {\bibfnamefont {M.}~\bibnamefont {Maiuri}}, \bibinfo {author} {\bibfnamefont {M.}~\bibnamefont {Garavelli}},\ and\ \bibinfo {author} {\bibfnamefont {G.}~\bibnamefont {Cerullo}},\ }\bibfield  {title} {\bibinfo {title} {Ultrafast spectroscopy: State of the art and open challenges},\ }\href@noop {} {\bibfield  {journal} {\bibinfo  {journal} {Journal of the American Chemical Society}\ }\textbf {\bibinfo {volume} {142}},\ \bibinfo {pages} {3} (\bibinfo {year} {2019})}\BibitemShut {NoStop}%
\bibitem [{\citenamefont {Kikuchi}(2015)}]{kikuchi2015fundamentals}%
  \BibitemOpen
  \bibfield  {author} {\bibinfo {author} {\bibfnamefont {K.}~\bibnamefont {Kikuchi}},\ }\bibfield  {title} {\bibinfo {title} {Fundamentals of coherent optical fiber communications},\ }\href@noop {} {\bibfield  {journal} {\bibinfo  {journal} {Journal of Lightwave Technology}\ }\textbf {\bibinfo {volume} {34}},\ \bibinfo {pages} {157} (\bibinfo {year} {2015})}\BibitemShut {NoStop}%
\bibitem [{\citenamefont {Suh}\ and\ \citenamefont {Vahala}(2018)}]{suh2018soliton}%
  \BibitemOpen
  \bibfield  {author} {\bibinfo {author} {\bibfnamefont {M.-G.}\ \bibnamefont {Suh}}\ and\ \bibinfo {author} {\bibfnamefont {K.~J.}\ \bibnamefont {Vahala}},\ }\bibfield  {title} {\bibinfo {title} {Soliton microcomb range measurement},\ }\href@noop {} {\bibfield  {journal} {\bibinfo  {journal} {Science}\ }\textbf {\bibinfo {volume} {359}},\ \bibinfo {pages} {884} (\bibinfo {year} {2018})}\BibitemShut {NoStop}%
\bibitem [{\citenamefont {Swann}\ and\ \citenamefont {Newbury}(2006)}]{swann2006frequency}%
  \BibitemOpen
  \bibfield  {author} {\bibinfo {author} {\bibfnamefont {W.~C.}\ \bibnamefont {Swann}}\ and\ \bibinfo {author} {\bibfnamefont {N.~R.}\ \bibnamefont {Newbury}},\ }\bibfield  {title} {\bibinfo {title} {Frequency-resolved coherent lidar using a femtosecond fiber laser},\ }\href@noop {} {\bibfield  {journal} {\bibinfo  {journal} {Optics Letters}\ }\textbf {\bibinfo {volume} {31}},\ \bibinfo {pages} {826} (\bibinfo {year} {2006})}\BibitemShut {NoStop}%
\bibitem [{\citenamefont {Black}(2001)}]{black2001introduction}%
  \BibitemOpen
  \bibfield  {author} {\bibinfo {author} {\bibfnamefont {E.~D.}\ \bibnamefont {Black}},\ }\bibfield  {title} {\bibinfo {title} {An introduction to pound--drever--hall laser frequency stabilization},\ }\href@noop {} {\bibfield  {journal} {\bibinfo  {journal} {American Journal of Physics}\ }\textbf {\bibinfo {volume} {69}},\ \bibinfo {pages} {79} (\bibinfo {year} {2001})}\BibitemShut {NoStop}%
\bibitem [{\citenamefont {Liang}\ \emph {et~al.}(2022)\citenamefont {Liang}, \citenamefont {Zhong}, \citenamefont {Tang}, \citenamefont {Liu}, \citenamefont {Yao}, \citenamefont {Sun}, \citenamefont {Zhang}, \citenamefont {Gao}, \citenamefont {Heidari}, \citenamefont {Qian} \emph {et~al.}}]{liang2022rotating}%
  \BibitemOpen
  \bibfield  {author} {\bibinfo {author} {\bibfnamefont {X.}~\bibnamefont {Liang}}, \bibinfo {author} {\bibfnamefont {Y.}~\bibnamefont {Zhong}}, \bibinfo {author} {\bibfnamefont {J.}~\bibnamefont {Tang}}, \bibinfo {author} {\bibfnamefont {Z.}~\bibnamefont {Liu}}, \bibinfo {author} {\bibfnamefont {P.}~\bibnamefont {Yao}}, \bibinfo {author} {\bibfnamefont {K.}~\bibnamefont {Sun}}, \bibinfo {author} {\bibfnamefont {Q.}~\bibnamefont {Zhang}}, \bibinfo {author} {\bibfnamefont {B.}~\bibnamefont {Gao}}, \bibinfo {author} {\bibfnamefont {H.}~\bibnamefont {Heidari}}, \bibinfo {author} {\bibfnamefont {H.}~\bibnamefont {Qian}}, \emph {et~al.},\ }\bibfield  {title} {\bibinfo {title} {Rotating neurons for all-analog implementation of cyclic reservoir computing},\ }\href@noop {} {\bibfield  {journal} {\bibinfo  {journal} {Nature Communications}\ }\textbf {\bibinfo {volume} {13}},\ \bibinfo {pages} {1549} (\bibinfo {year} {2022})}\BibitemShut {NoStop}%
\bibitem [{Standard Performance Evaluation Corporation()}]{spec}%
  \BibitemOpen
  Standard Performance Evaluation Corporation,\ \href@noop {} {\bibinfo {title} {Spec benchmarks and tools}},\ \bibinfo {howpublished} {\url{https://www.spec.org/benchmarks.html}} (\bibinfo {year} {2024}),\ \bibinfo {note} {accessed: 2024-12-31}\BibitemShut {NoStop}%
\bibitem [{PassMark Software()}]{passmark}%
  \BibitemOpen
  PassMark Software,\ \href@noop {} {\bibinfo {title} {Passmark - cpu benchmarks}},\ \bibinfo {howpublished} {\url{https://www.cpubenchmark.net/CPU_mega_page.html}} (\bibinfo {year} {2024}),\ \bibinfo {note} {accessed: 2024-12-31}\BibitemShut {NoStop}%
\bibitem [{HWBOT()}]{hwbot}%
  \BibitemOpen
  HWBOT,\ \href@noop {} {\bibinfo {title} {elmor`s cpu frequency score 9117.75 mhz with core i9 14900ks (8p)}},\ \bibinfo {howpublished} {\url{https://hwbot.org/submission/5508265_elmor_cpu_frequency_core_i9_14900ks_(8p)_9117.75_mhz}} (\bibinfo {year} {2024}),\ \bibinfo {note} {accessed: 2024-12-31}\BibitemShut {NoStop}%
\end{thebibliography}%
\end{document}



\title{\Large{Supplementary Information for ``All-optical computing with beyond 100-GHz clock rates''}}

\author{Gordon H.Y. Li$^{1,*}$, Midya Parto$^{2,3,4,*}$, Jinhao Ge$^{5,*}$, Qing-Xin Ji$^{5}$, Maodong Gao$^{3, 5}$, Yan Yu$^{5}$, James Williams$^{2}$, Robert M. Gray$^{2}$, Christian R. Leefmans$^{1}$, Nicolas Englebert$^{2}$, Kerry J. Vahala$^{5}$, and Alireza Marandi$^{1,2,\dagger}$\\
\small{$^{1}$Department of Applied Physics, California Institute of Technology, Pasadena, CA 91125, USA\\
$^{2}$Department of Electrical Engineering, California Institute of Technology, Pasadena, CA 91125, USA\\
$^{3}$Physics and Informatics Laboratories, NTT Research, Inc., Sunnyvale, California 94085, USA\\
$^{4}$CREOL, The College of Optics and Photonics, University of Central Florida, Orlando, FL, USA\\
$^{5}$T. J. Watson Laboratory of Applied Physics, California Institute of Technology, Pasadena, California 91125, USA\\
$^{*}$These authors contributed equally\\
$^{\dagger}$marandi@caltech.edu\\}}
\maketitle

\newpage
\setcounter{figure}{0}
\renewcommand{\thefigure}{S\arabic{figure}}
\section{Experimental setup}
\begin{figure}[h]
\includegraphics[width=\linewidth]{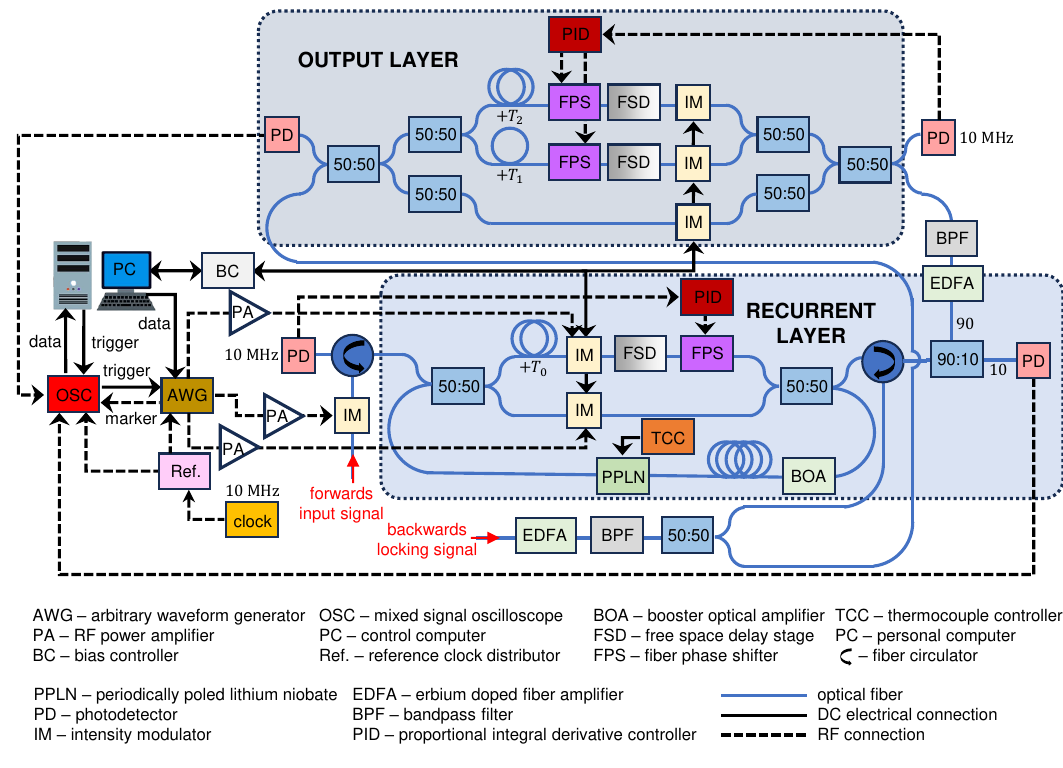}
\caption{\textbf{Detailed schematic of AO-RNN experimental setup.}}
\label{fig:S1}
\end{figure}
\newpage
\section{Noisy waveforms}
\begin{figure}[h]
\includegraphics[width=\linewidth]{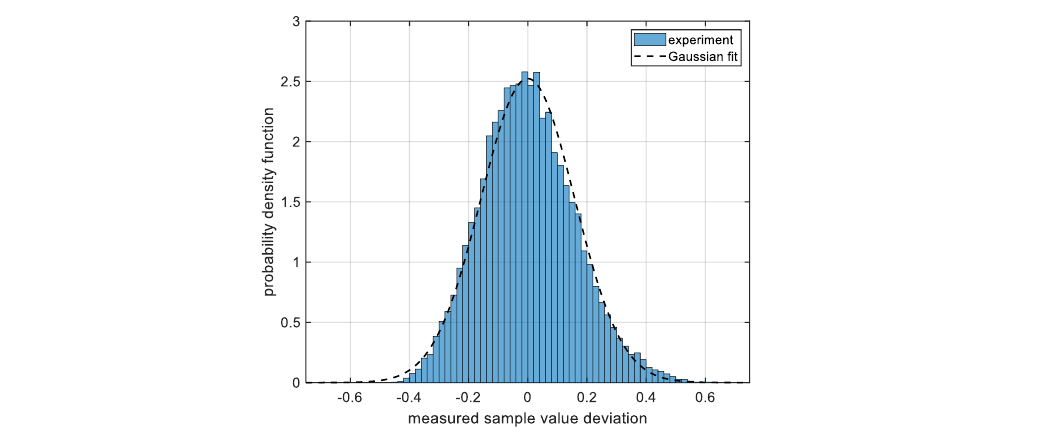}
\caption{\textbf{Noise distribution for noisy waveform input signals.} The histogram shows the statistics for the deviation between the ideal noiseless waveform samples and the measured optical input signal values (amplitudes normalized to [-1,1]) for 8000 input waveform sample points. It is well-approximated by a Gaussian distribution (dashed black line) with zero mean and standard deviation of $\sim0.158$.}
\label{fig:S2}
\end{figure}

\section{Electro-optic frequency comb}
\begin{figure}[b]
\includegraphics[width=\linewidth]{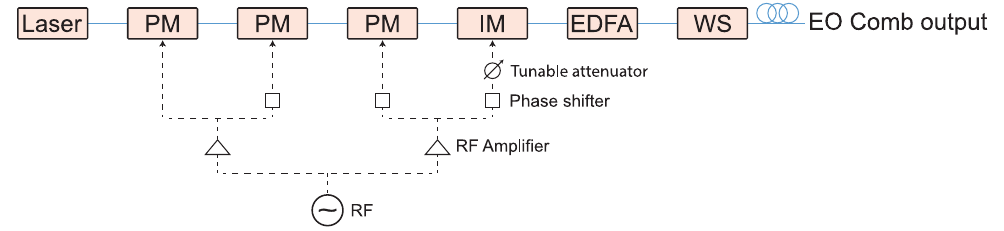}
\caption{\textbf{Experimental setup for Electro-optic frequency comb.}
Abbreviations: PM, phase modulator; IM, intensity modulator; EDFA, erbium-doped fiber amplifier; WS, waveshaper.}
\label{eocomb_setup}
\end{figure}

\begin{figure}[t]
\includegraphics[width=\linewidth]{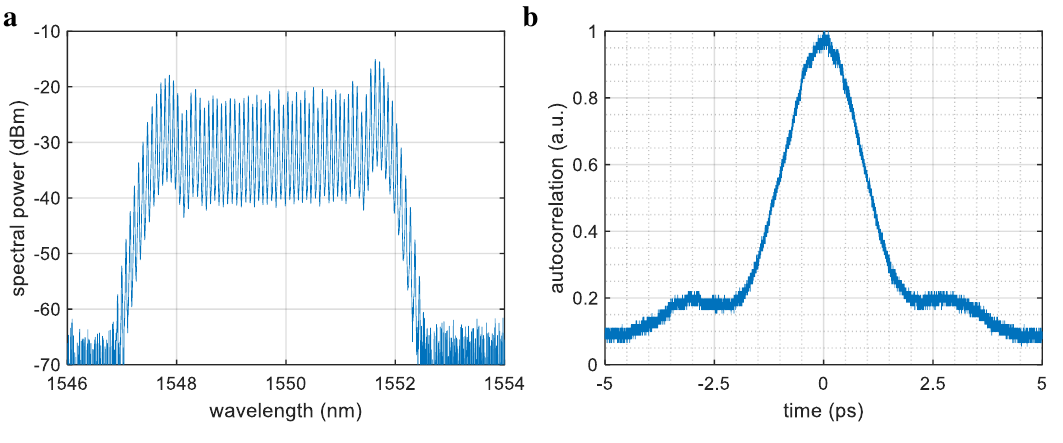}
\caption{\textbf{Electro-optic frequency comb characterization.} (a) Optical spectrum and (b) autocorrelation for the electro-optic frequency comb operating with a repetition rate of $10~\mathrm{GHz}$.}
\label{fig:S4}
\end{figure}

The physical configuration of the electro-optic frequency comb system is illustrated in Fig.~\ref{eocomb_setup}. A continuous-wave, single-frequency laser is phase-modulated by three cascaded modulators, generating a series of sidebands that form a frequency comb, as shown in Fig.~\ref{fig:S4}(a). The spacing between adjacent comb lines is determined by the radio-frequency (RF) signal driving the phase modulators, which is set to \textcolor{black}{10} GHz in our experiment. To maximize the number of comb lines, the RF signals driving the individual modulators are amplified and phase-shifted to ensure effective in-phase modulation. Following phase modulation, an additional intensity modulation stage is introduced to flatten the comb spectrum. The intensity modulator (IM) is biased at the half-power point of its transmission curve, and the modulation signal frequency matches that of the phase modulators. The modulation strength is controlled using a variable RF attenuator. Subsequently, the comb is amplified using an erbium-doped fiber amplifier (EDFA) to facilitate further operations with enough power. To generate optical pulses in the time domain, the phase of each comb line is adjusted using a programmable waveshaper, which applies second-order dispersion compensation to the frequency comb. After phase compensation, the time-domain field becomes a series of optical pulses, characterized by an autocorrelator, as shown in Fig.~\ref{fig:S4}(b). The dispersion applied by the waveshaper is optimized by minimizing the pulse width measured through autocorrelation. 


\section{Soliton microcomb}
\begin{figure}[h]
\includegraphics[width=\linewidth]{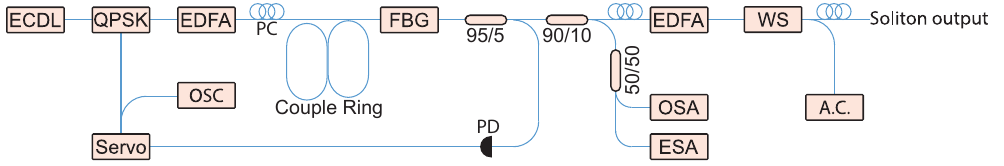}
\caption{\textbf{Experimental setup for generating the multi-soliton state in coupled-ring resonators.}
Abbreviations: ECDL, external cavity diode laser; QPSK, quadrature phase shift keying; EDFA, erbium-doped fiber amplifier; FBG, fiber Bragg grating; PC, polarization controller; PD, photodetector; OSA, optical spectrum analyzer; ESA, electrical spectrum analyzer; WS, waveshaper; OSC, oscilloscope; A.C., autocorrelator.}
\label{soliton_setup}
\end{figure}

\begin{figure}[h]
\includegraphics[width=\linewidth]{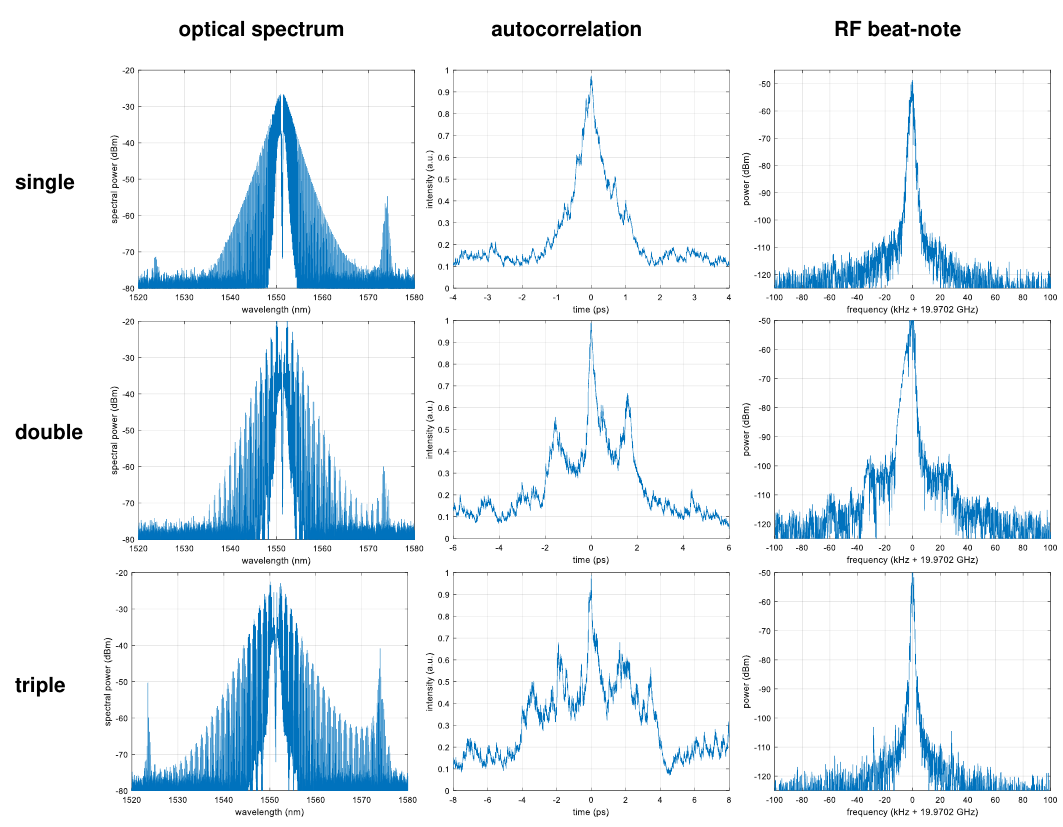}
\caption{\textbf{Bipartite-soliton pulse pair state characterization.} The optical spectrum (left column), autocorrelation (middle column), and RF beat-note (right column) for the single (top row), double (middle row), and triple (bottom row) bipartite-soliton pulse states.}
\label{fig:S6}
\end{figure}

The coupled-ring device is fabricated using an ultra-low-loss Si$_3$N$_4$ platform~\cite{jin2021hertz}. It consists of two partially coupled racetrack resonators with slightly different free-spectral ranges (FSRs), which provide anomalous dispersion required for bright soliton generation. The laser output is modulated by a fast single-sideband modulator (QPSK) and subsequently amplified by an erbium-doped fiber amplifier (EDFA). The QPSK modulator facilitates the rapid frequency sweeping of the pump laser in order to avoid thermal effect during the soliton locking process~\cite{yi2015soliton}. Lens fibers (not shown in Fig. ~\ref{soliton_setup}) are used to couple light in and out of the coupled-ring system. The through-port output is filtered using a fiber Bragg grating (FBG) to separate the comb and pump signals. The filtered signal is then split into multiple beams for different purposes. One beam is directed to a photodetector, which measures the comb power that are used for stabilizing the pump-resonance detuning. Another beam is sent to an optical spectrum analyzer (OSA) and an electrical spectrum analyzer (ESA) for characterizing soliton spectrum and RF beat-note signal. The remaining beam, carrying the main comb power, is amplified by an EDFA and then passed through a waveshaper to compensate the fiber dispersion. The amplified soliton is then directed to the AO-RNN (all-optical recurrent neural network) setup.

When single, double, or triple soliton pulse pairs are formed in the resonator, the measured total comb power varies accordingly. By selectively stabilizing specific comb power levels, different stable soliton states can be reliably and individually locked for classification experiments, as shown in Fig.~\ref{fig:S6}. For comparative analysis, the multi-soliton states are further characterized and validated using optical spectrum measurements and autocorrelation signals. These characterization results provide a detailed basis for the classification experiments conducted with the AO-RNN setup. The soliton spectrum and autocorrelation traces for different pulse numbers are shown in Fig. \ref{fig:S6}. The soliton spectrum exhibits an approximately $sech^{2}$ envelope shape, with dispersive waves~\cite{yang2016spatial} appearing at frequencies where the mode and comb frequencies coincide, leading to resonant power enhancement. The soliton repetition rate, measured using the ESA, is approximately 19.97 GHz with a resolution bandwidth of 1 kHz. The autocorrelation of the generated periodic soliton pulse train is measured using an autocorrelator, and the result is shown in Fig.~\ref{fig:S6}. For single, double, and triple soliton states, the autocorrelation traces display 1, 3, and 5 peaks, respectively. As an additional note, the formation location of solitons within the resonator is a random process and is not actively controlled during the experiment (only the number of solitons is controlled). The data presented in Fig. \ref{fig:S6} are selected as representative results.

\section{Ultrafast optical input time-multiplexing}
\begin{figure}[b]
\includegraphics[width=\linewidth]{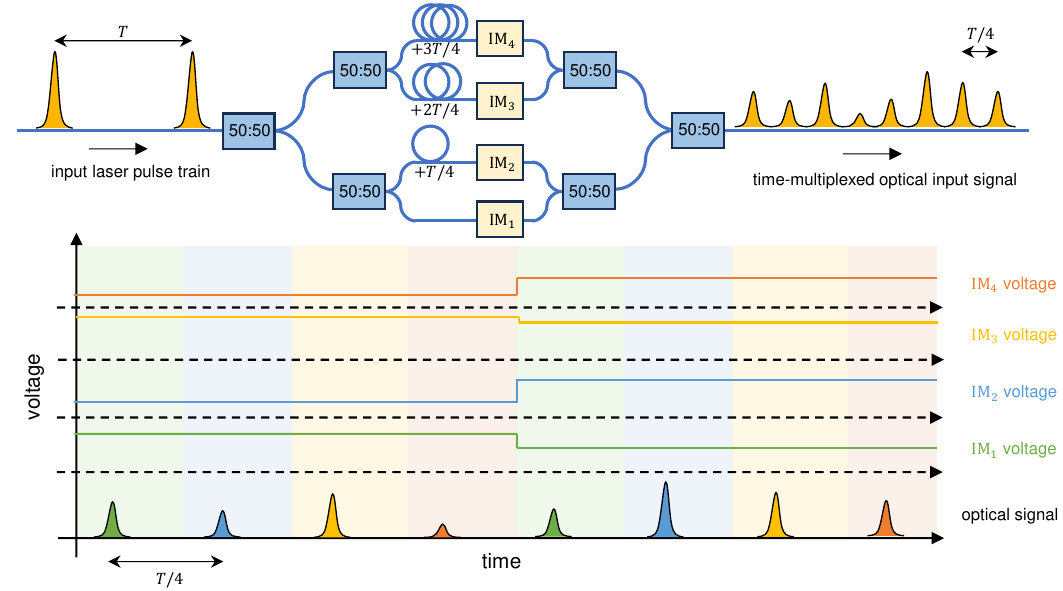}
\caption{\textbf{Real-time time-multiplexed optical input generation.} A multi-arm Mach-Zehnder interferometer in which each arm is delayed by an integer fraction of $T$ relative to the other arms and contains an intensity modulator (IM) with sampling period of $T$ can generate an optical input signal with effective sampling period that is an integer fraction of $T$.}
\label{fig:S7}
\end{figure}
The signal information in the AO-RNN is encoded onto the coherent amplitude of ultrashort laser pulses. One unique advantage of this approach over using continuous-wave light is that it allows for optical time-interleaving techniques to multiply the effective repetition rate and input sampling rate. For the task of noisy waveform classification, we can use two different time-multiplexing techniques to generate equivalent ultrafast optical input signals beyond the limited sampling rate of our arbitrary waveform generator. The first technique allows for real-time input generation as shown conceptually in Fig.~\ref{fig:S7}. Consider an input laser pulse train with repetition period $T$. Suppose we wish to increase the effective input sampling rate by a factor of 4. Then, to do this, we use a Mach-Zehnder interferometer with 4 arms. Each arm is delayed by $T/4$ relative to the previous arm and contains an intensity modulator with sampling period of $T$. Upon recombining at the output, the optical input signal effectively has an input sampling period of $T/4$. This method exploits the fact that the pulse length of the laser pulse $\tau$ is much less than the RF sampling period $T$. The maximum allowable number of arms in the Mach-Zehnder interferometer to upconvert the sampling rate is $\sim T/\tau$ before the output optical pulses begin to undesirably overlap temporally. 

\begin{figure}[h]
\includegraphics[width=\linewidth]{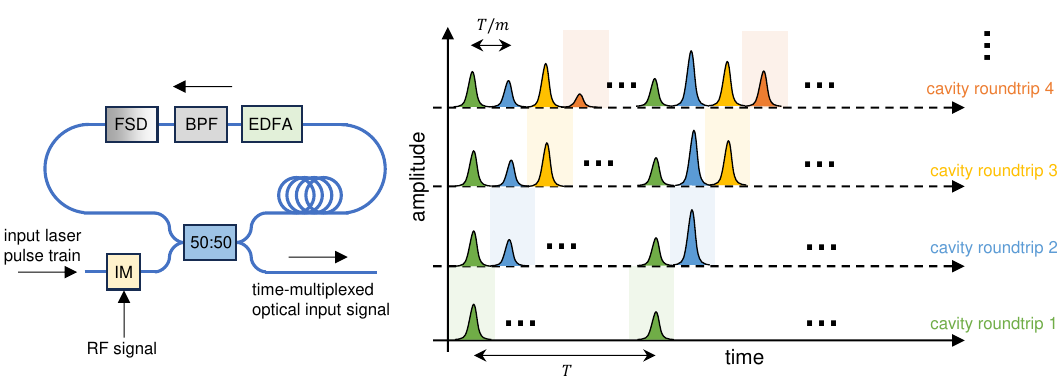}
\caption{\textbf{Offline time-multiplexed optical input generation.} The desired optical input signal is gradually built-up over many roundtrips of an asynchronously-pumped optical cavity. IM: intensity modulator, EDFA: erbium doped fibre amplifier, BPF: band-pass filter, FSD: free-space delay.}
\label{fig:S8}
\end{figure}

In practice, we are limited by the number of available modulators and channels on our arbitrary waveform generator. The scope of this work is to demonstrate ultrafast optical computing, and not to design new ultrafast optical transceivers. Therefore, to reach even higher effective input sampling rates for very high clock rate computing, we use another time-multiplexing technique based on an asynchronously-pumped optical cavity. As opposed to real-time input generation, we call this technique ``offline'' input generation since the input signals must be prepared ahead-of-time before beginning the AO-RNN computation. The concept for offline input generation is shown in Fig.~\ref{fig:S8}. Suppose that we have an input laser pulse train with repetition period $T$ and we can generate RF samples with period $T$, but desire to have an input optical signal with sample period of $T/m$ where $m>1$. A synchronously-pumped optical cavity has a roundtrip time $NT$ where $N$ is an integer so that each input laser pulse will overlap temporally with a laser pulse in the cavity. We detune the cavity by adjusting a free-space delay stage such that the roundtrip time is reduced by $T/m$, so that the rountrip time of the asynchronously-pumped optical cavity is $NT-T/m$. During the first cavity roundtrip we modulate the input laser pulses with every $m^{\mathrm{th}}$ waveform sample (i.e. sample 1, $m+1$, $2m+1$, $\ldots$). Then, during the next cavity roundtrip we modulate the input laser pulses with every $m^{\mathrm{th}}$ waveform sample offset by 1 sample (i.e. sample 2, $m+2$, $2m+2$, $\ldots$). The cavity pulses from the first roundtrip will be displaced forwards by $T/m$ relative to the input laser pulses for the second roundtrip. We repeat this procedure until we have completed modulation of all waveform samples. In this way, we gradually build-up the desired optical waveform with sampling period $T/m$ over multiple cavity roundtrips using only a single input modulator and RF channel with sampling period $T$. The cavity contains an EDFA that is tuned to compensate the roundtrip loss so that many (typically $>20$) cavity roundtrips are possible without exponential signal degradation from the cavity out-coupling and propagation loss. Amplitude scaling factors to account for loss variations between roundtrips can be calibrated by sending a single laser pulse into the cavity and measuring its amplitude decay after each roundtrip. Using this offline input generation method, the cavity roundtrip time must be at least $>nT/m$ where $n$ is the total number of waveform samples. The maximum upconversion rate factor is limited by $\tau\approx T/m$ where $\tau$ is the input laser pulse length. If the cavity detuning $T/m$ is comparable to the pulse length $\tau$, then pulses will overlap from roundtrip to roundtrip, which will result in undesirable sample cross-talk. The gating modulator controlling inputs into the AO-RNN is synchronized to the offline input generation such that it only allows the final desired cavity roundtrip to be transmitted, which prevents input signal artifacts from the previous roundtrips.   

\section{All-optical image generation}
\begin{figure}[h]
\includegraphics[width=\linewidth]{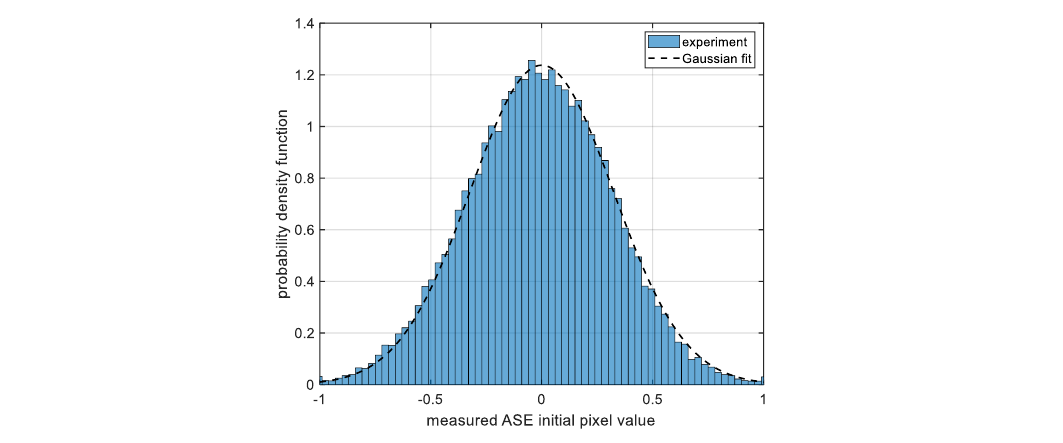}
\caption{\textbf{Quantum noise distribution.} The histogram shows the statistics for the initial pixel values sampled from amplified spontaneous emission (ASE) for $\sim35000$ sample points. It is well-approximated by a Gaussian distribution (dashed black line) with zero mean and standard deviation of $\sim0.322$.}
\label{fig:S9}
\end{figure}

\begin{figure}[h]
\includegraphics[width=\linewidth]{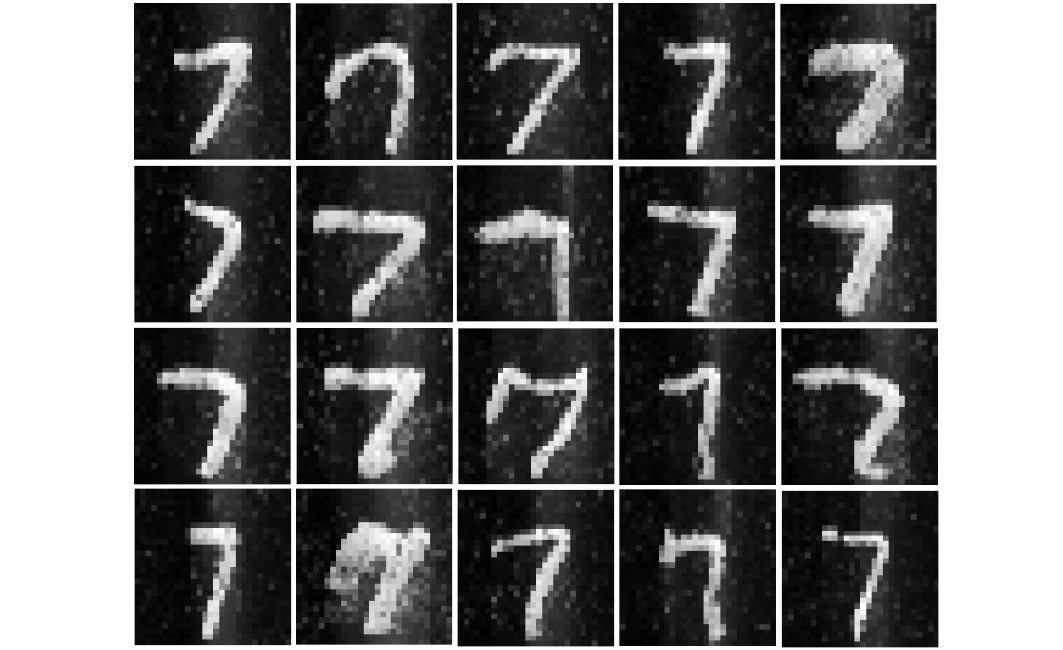}
\caption{\textbf{25 generated images of ``seven'' using the AO-RNN.}}
\label{fig:S10}
\end{figure}

\bibliography{supp_refs}